# AI Risk Skepticism – A Comprehensive Survey

**Vemir Michael Ambartsoumean**
AI Safety, Policy, Ethics
Independent Researcher
vemireal@gmail.com

**Roman V. Yampolskiy**
Computer Science and Engineering
University of Louisville
roman.yampolskiy@louisville.edu

## Abstract

In this thorough study, we took a closer look at the skepticism that has arisen with respect to potential dangers associated with artificial intelligence – denoted as AI Risk Skepticism. Our study takes into account different points of view on the topic and draws parallels with other forms of skepticism that have shown up in science. We categorize the various skepticisms regarding the dangers of AI by the type of mistaken thinking involved. We hope this will be of interest and value to AI researchers concerned about the future of AI and the risks that it may pose. The issues of skepticism and risk in AI are decidedly important and require serious consideration. By addressing these issues with the rigor and precision of scientific research, we hope to better understand the objections we face and to find satisfactory ways to resolve them.

## Introduction

Artificial intelligence, with its rapidly advancing capabilities, presents many potential perils to humanity. Given the likely benefits of advanced AI, it is no surprise that skepticism has emerged about the dangers. In this survey, we delve deeper into this skepticism by exploring different schools of thought and drawing parallels with other forms of skepticism in science. To begin our examination, we will classify the different species of skepticism around the risk of AI – a crucial step in our investigation – as it will allow us to analyze the root causes of these doubts with greater precision. Through this process, we seek to uncover the many shortcomings in these arguments, implicitly reveal the motivations of these critics, and provide informed responses. The discourse surrounding skepticism about the risks of AI is of the utmost importance – and demands our careful attention and consideration.

We will begin by explaining the difference between AI Risk Denialism and AI Risk Skepticism. In the field of AI Risk, the distinction between AI Risk Denialism and AI Risk Skepticism is essential. The former represents a disregard for the dangers posed by AI, dismissing the idea that such dangers even exist, while the latter embodies a more measured and discerning perspective, one that recognizes the potential for harm but remains skeptical of its extent. AI Risk Denialists maintain that AI holds great promise for humanity and that the benefits of this technology far outweigh any potential risks. They may argue that the possibility of AI causing harm is negligible and that rapid advances in AI will effectively reduce these risks. In contrast, AI Risk Skeptics embrace a more respectful and critical view of AI's potential. Acknowledging that AI does indeed pose some risks, they remain skeptical of the full extent of these risks and advocate for further research to better understand them and develop effective strategies for integration. It therefore becomes clear that AI Risk Denialism represents a public dismissal of the risks raised by AI, while AI Risk Skepticism embodies a more valued and informed perspective, one that recognizes the potential for harm but remains grounded in critical inquiry. Arguments of various forms that are skeptical, and in complete denial of, AI's associated risks are represented in this study, and deconstructed appropriately.

## Main Categories

### *Crux of Time Illusion*

*This section concentrates on the oft-purported claim by skeptics that AI is not risky now, that future innovations and breakthroughs in this field are so far off that there is no need or ability to worry now, or some other manipulation of time is used to assure a reader against combating skepticism seriously. Often there is confidence in our ability to adapt to change as we have before, though this is constantly misplaced, as the reader will see where AI, in its exponential and unpredictable pattern of progress, will be more and more difficult to keep pace with and maintain any safety standards at all without a significant and conscious effort.*

*As will be emphasized in a future section – **Using History to Predict the Unpredictable Future** – using the past to predict the future of AI in favor of skepticism of risk stands on shaky ground. Blind progress will also be seen in this paper as counterintuitive in the long-term perspective of AI as a benefit to the world and humanity overall.*

A frequent argument against work on AI Safety is that we are hundreds, if not thousands, of years away from developing Superintelligent machines, and therefore – even if they may present some danger, it is a waste of human and computational resources to allocate any effort to address Superintelligence Risk at this point in time – emphasizing that there are still scientific leaps to make [1], that evolution induced a large amount of computation difficult to replicate [2], and that we are far further away [3] [4] [5-7][1] than some "pundits" might claim [8]. Overall, there is the feeling that we have plenty of time to prepare[2] [3] [2, 9, 10], and that the issue, or certain aspects of the issue, are too far away to be worth researching [11]. Researchers such as Andrew Ng, who work on current cutting-edge systems, dismiss future concerns about advanced AI, that the "future is so uncertain" [12] that such concerns cannot be meaningfully considered (ibid.), prioritizing immediate [13] and more clearly defined problems [12]. A lack of imagination cannot be a prerequisite for ignoring long-term risks, as issues will arise regardless of a certainty of foresight or perspective.

Such a position as aforementioned also doesn't properly take into account the possibility that it may take even longer to develop appropriate AI Safety mechanisms, therefore the perceived abundance of time[4] is a feature, not a bug. It also ignores a non-zero possibility of an earlier development of Superintelligence – indeed, the logic falls, as with other global risks, towards preparing immediately – as we risk the definite possibility of being too *late*. There is also a wide range of technical problems to be addressed in the present moment that are both relevant for near-term and long-term AI, making the distinction or bifurcation of focus unnecessary [12]. Awareness of this point alone would reduce division in the field, as there would be no need for delay in considering the conglomerate of these issues [15] – as concern is not justifiably predicated on imminence [16]. 'Hard to predict' is not equivalent to 'impossible to prepare for.' There is also false confidence in relation to many variables within the field, such as how slowly AI will progress (with historic reliance on how we tend to overestimate capabilities and underestimate difficulties [3]), being able to adjust along the way with developing technologies, and being able to work out the 'kinks' without sudden and unexpected issues [8]. This is a very near-term issue, in which the lack of urgency simply lies in a lack of education [17], and the idea that we should "wait for the canaries to collapse" [18] or "fear it when we get to it," is an unfortunate, illusory luxury that will backfire accordingly, akin to purchasing a fire extinguisher when your house is already aflame.

---

[1] This source emphasizes life extension and any predictions at all as too far off. 7. Spectrum, I. *HUMAN-LEVEL AI IS RIGHT AROUND THE CORNER—OR HUNDREDS OF YEARS AWAY*. IEEE Spectrum 2017; Available from: https://spectrum.ieee.org/humanlevel-ai-is-right-around-the-corner-or-hundreds-of-years-away.
[2] Or, that there has not been a catastrophe of this magnitude yet, as an argument for a lack of a future one.
[3] Rohin Shah asserts that we have a 90% likelihood of things going well *by default*. He also feels that slow takeoff is about graduality rather than calendar time, giving us space to figure problems out. 2. Asya Bergal, R.L., Sara Haxhia, Rohin Shah, *Conversation with Rohin Shah*, in *AI Impacts*. 2019.

One response to a common argument – "'It's like worrying about overpopulation on Mars.' This is an interesting variation on 'too soon to worry,' one that appeals to a convenient analogy: not only is the risk easily managed and far in the future, but also it is extremely unlikely we would even *try* to move billions of humans to Mars in the first place. The analogy is a false one, however. We *are already* devoting huge scientific and technical resources to creating ever-more-capable AI systems. A more apt analogy would be a plan to move the human race to Mars with no consideration for what we might breathe, drink, or eat once we arrive." [16]

A mention of the past to predict future time scales of growth – "The development of what we might call massive superhuman intelligence will be an evolutionary process involving changes in the social, physical, and intellectual fabric on which our society is built. Changes like that take time[5]" [19]. This is true (aside from the assumption embedded in the subtext of a slow time scale), however it leaves no excuse for early preparation and laying fertile ground for exponential technologies – which would also assist in the noted evolutionary process and its components. Also, soft takeoff could be in the order of ten or twenty years, which is a short time in practical terms considering the multi-sectoral affect [20] and long-term effect of these innovations.

One assumption is that we will be able to correct the system online – "AI systems are likely to fail in not extinction-level ways before they fail in extinction-level ways, and presumably we will learn from that and not just hack around it and fix it and redeploy it… someone noticed that AI systems are not particularly fair and now there's just a ton of research into fairness" [2] – essentially stating that the pros of AI will outweigh the cons (including unfairness) (ibid.), and we can handle the emerging issues without great distress. This is not necessarily true. While no researcher or engineer knows the actual take-off speed at this point in time, or how a system

---

[4] –and difficulty. 14.    Mitchell, M., *Why AI is Harder Than We Think.* 2021.
[5] Michael Littman also believes that the idea of recursive self-improvement is far from the core direction of AI, which downplays the amount of safety needed. 19.    Littman, M., *Interview with Michael Littman on AI risks*, A. Kruel, Editor. 2015.

might fail in the future, it is prudent to be ready for the worst-case scenario. Indeed, the ethics of smart machines may be a "slow inquiry…be subject to revision," [21] time we likely do not have, with a need to proceed immediately with great effort. It seems more clear that waiting for an issue to arise while innovating simply in favor of the 'pros' can only be effective, escaping danger, and non-hazardous for so long, and this ignores the difficulty of preventing or mitigating a runaway system or facing the (mounting current) effects of AI's 'cons.[6]'

We must have a holistic process – "Any progress that doesn't contribute to the solution but brings forward the date by which we *must* solve it…is bad" [22]. Essentially, moving forward with AI progress and immediate benefits without tackling the big picture – hard problems in AI Safety – such as value alignment, is harmful. Also, simply because AI hard takeoff is hard to imagine [2], as is most exponential growth, does not reduce its plausibility or inherent danger.

## *We Are Unsure About Something, Thus Don't Worry*

*AI risk skepticism has a common thread within the arguments of this section, making a mutual critical error – that because we do not fully understand a particular feature about humans or machines (intelligence, certain capabilities), that we need not worry. That these holes would reduce any responsibility, or that particulars are not obvious at present, are weak arguments diverting away from a field, AI Safety, that would likely create sustainable progress far into the future, at minimum providing us a better chance in doing so. A lack of understanding is not a proper foundation to avoid exploring how we may best approach these solutions*

*In a further section – "**Something is Impossible Illusion**" – the reader will see also that many skeptics claim something is completely impossible in the realm of innovation, and therefore preparation is certainly unnecessary. This confidence is again a simple yet critical mistake, as many things in AI that exist today would have been considered impossible one generation ago, and many risks that exist and impact humanity today may have been avoided with a bit of foresight in considering possibility.*

There is continued doubt in our progress towards generality, every step of the way – "[t]asks like chess playing have been solved using the exponential advance in brute force rather than general cognition" [21] – and much of the disagreement still exists because there are still not only power [23] and other limitations, but a concrete pathway to discover (in other words, finding "miracles of understanding" [21]). For example – in deep learning, uncertainty can or will be reduced with new solutions – "expert disagreement provides a road map for progress towards artificial intelligence" [24]. Erik Larson argues in kind that the hard work needed to be done for progress in AGI is not being done, in favor of more practical applications, and therefore there is no clear pathway at present [25][7]. Arguments often arise that the sci-fi-like horrors are only grounded in an eventuality (or the opposite) that has no clear path from current tools, and all are quite far off [26]. Utility functions are accused of being too narrow, unlikely to get us to AGI, and future implementations will be very different from present approaches, making "working on AGI" [27] impossible now (ibid.). Often the argument is to simply wait until we have such strong models [23]. While we are making good progress on AI, it is not obvious how to get from our current state in AI to AGI, [28][8] [11][9] [29] and current methods and narrow solutions may not scale [25, 29-31][10]. This may be true, however – we certainly need all the time available to develop necessary safety mechanisms. For example, companies require a sophisticated methodology to look out for customer wellbeing, and without any alignment techniques in place, this consideration-to-implementation will not exist, and companies will simply not bother [23]. Additionally, current state-of-the-art systems do not seem to hit limits yet, subject to availability of compute for increasing model size. At the pace of advancement, alignment research is needed *in advance* and is not bottlenecked by scale of training – one subject example being the improvement of human preferences methodology[11] [23].

---

[6] 'Pros' and 'cons' are labeled here in such a way with respect to the amount of overlap and a gray area that exists when discussing AI's capabilities and impact upon humanity.

[7] He also references abductive inference as necessary to produce AGI, and that we don't have the ability to arrive there with our current hypotheses on the state of affairs in computer science. 25. Dembski, W.A. *Artificial Intelligence: Unseating the Inevitability Narrative*. 2021; Available from: https://mindmatters.ai/2021/04/artificial-intelligence-unseating-the-inevitability-narrative/.

One argument considers many if-then contingencies for AI to become an existential risk [30], "a failure of which at any point would negate the apocalypse" [9]. This is incorrect. AI can be dangerous, and is harmful even at its current state, and it is evident that danger only compounds with more complexity and sophistication of the machine. To emphasize both our responsibility as humans, and the danger of a completely benign system – the right machine, in the wrong hands, is a disaster.

"We didn't rush to put rules in place about how airplanes should work before we figured out how they'd fly in the first place[12]" [32]. Some people may agree with concerns about Superintelligence, but argue that AI Safety work is not possible or futile in the absence of a Superintelligent AI[13] on which to run experiments or with which to have context [19][14]. This view is contradicted by a significant number of publications produced by the AI Safety community in recent years, and the co-author of this paper in particular. Also, the analogy is not a direct one, as we already have AI which engages with the world today – making both innovation and safety considerations to be an ongoing spectrum rather than a binary-like, backloaded engagement.

---

[8] This source specifically doubts our understanding and extrapolation of intelligence. 28. Kaufman, J. *Conversation with Michael Littman*. 2017; Available from: https://www.jefftk.com/p/conversation-with-michael-littman.

[9] Rodney Brooks states here, "I think the worry stems from a fundamental error in not distinguishing the difference between the very real recent advances in a particular aspect of AI and the enormity and complexity of building sentient volitional intelligence," marking a gap from narrow to general AI. 11. *Existential risk from artificial general intelligence*, in *Wikipedia*. Wikipedia.

[10] This reference also indicates that being cautious and prepared about this issue is important, but not obsessive. 30. Winfield, A., *Artificial intelligence will not turn into a Frankenstein's monster*, in *The Guardian*. 2014.

[11] This can be used as an example with current systems, and there are many references and resources regarding alignment that strengthen this claim.

[12] This is in tandem with a fear of losing out on gains or 'pros.'

[13] –or a more complete, deeper understanding of intelligence. This is one main takeaway (and emphasis) made by Littman 19. Littman, M., *Interview with Michael Littman on AI risks*, A. Kruel, Editor. 2015.

[14] Michael Littman: "I think the idea of seriously studying AGI safety in the absence of an understanding of AGI is futile. At a high level, raising awareness and scoping out possibilities is fine. But, proposing specific mechanisms for combatting this amorphous threat is a bit like trying to engineer airbags before we've thought of the idea of cars. Safety has to be addressed in context and the context we're talking about is still absurdly speculative." 19. ibid.

One argument states that because we do not fully understand consciousness or morality, it is impossible to make predictions for AI and what it will possess. Therefore, AI Safety research is purported by skeptics as a waste of time [33]. Shying away from any research (and making broad claims about what is truly achievable) seems to be the least effective method in finding a solution. Many of the broader questions that require answering, such as ethical concerns, morality, and consciousness, are a concrete part of AI Safety research[15], and overlap with the unknown does not exclude more present risks that do not involve these accused-to-be ephemeral subjects. [16]

There is disagreement about the ability to formalize intelligence and rationality, that there may be no platonic idea of intelligence to find and code – "intelligence is just like a broad… just like conglomerate of a bunch of different heuristics that are all task specific, and you can't just take one and apply it on the other space. It is just messy and complicated and doesn't have a nice crisp formalization" [2]. Though we cannot say with certainty, the crux of heuristics determining and forming intelligence, especially when looking at higher forms of intelligence and rational capability – may be an appearance of causality, and formalizing rationality may be the result of more advanced research and development. Humans and AGI may come from different cores, or cannot be surmised as a conglomerate of nearby heuristics. This discovery may even be more of a risky outcome as it will be even farther along the scale of unpredictability due to uncanniness. Even taking the original argument at face value, a machine with a mastery of our embedded heuristics is still clearly dangerous.

"Current AI systems are relatively simple mathematical objects trained on massive amounts of data, and most avenues for improvement look like just adding more data. This doesn't seem like a recipe for recursive self-improvement." [34] This does not robustly address the issue long-term, as new possibilities could immediately create a self-improving system, thus creating a problem immediately without a solution in place. There is also relevant technical research to prepare for this situation right now [34].

The accusation that there is no one specific feature to optimize for self-improvement [35] can be argued against when looking at optimization strategy and expected goal utility maximization, among other secondary ways that an AI system can improve itself generally and in achieving its outcomes.

## *Something Else Takes Priority*

---

[15] Another pointed response in the article – "It seems that – worst case scenario – we're laying some foundational ideas for the 'possibility space' of problems managing AI, and the constellation of possible responses and approaches to deal with these challenges. Maybe when AGI is on the horizon, it'll be drastically different than humans imagine now, and all this thinking will be for naught, but it seems that – at least on some level – it's worth a shot. At least some of it." 33.    Fagella, D. *Is AI Safety Research a Waste of Time?* 2018; Creating or Enhancing Consciousness, Reflecting on What I've Read, The Good Must Be Explored]. Available from: https://danfaggella.com/is-ai-safety-research-a-waste-of-time/.

[16] It could be emphasized here that a plan of action and evergreen strategies can provide a better pathway to success in discovering the unknown subjects mentioned, in a safer manner. Also, it cannot possibly be a significant 'scheme' to collect money – if all AI Safety researchers lost their jobs (which is a comparatively strong minority), because we have discovered that safety mechanisms are easier to compile than previously thought (which is unlikely), then humanity is in a better position than to never have made the consideration at all.

*There seems to be a false appearance of scarcity in this discourse – in that considerations of AI Safety, or some considerations within the larger field of AI itself, need to be placed aside in favor of issues that skeptics feel are more important or require a higher priority – especially in regards to more abstract issues (eg. AGI, Conscious AGI). As the reader may observe, AI Safety is in no way an either-or issue or subject, and emphasis on building safe machines presents a way for many of these mentioned "prioritized" issues to also be resolved (potentially more effectively so). Attention to risk itself is valuable, forward-thinking, and can bring new awareness to emerging subfields when preparing for a beneficial future.*

*It should be mentioned, especially in this section, that we cannot in good faith comment as to whether any argument by a skeptic has an ulterior motive or hidden agenda, but it is suggested that the reader explore this concept on both sides of AI risk as to why an otherwise weak argument would be so stringently purported, evidenced or extrapolated by its subtext.*

Some proponents claim that immediate issues with today's AIs, such as unethical robots [30], mistakes with self-driving cars [31], algorithmic and data bias [36-38], privacy concerns [6], technological unemployment[17] [6, 26, 31, 39] and livelihood affect – including the distribution of benefits and risks [40], various public and specific concerns within select communities [40], limited transparency, surveillance [39], among many others[18] should take precedence over concerns about future technology (ie. AGI/Superintelligence), which does not yet exist and may not exist for decades [41],[19] especially when facing the danger of current military applications [6]. This topic is not an either-or issue, and in response there have been several olive branches extended [41] to support both near- and long-term AI Safety considerations. Setting evergreen perspectives in this field will allow the inevitably more powerful technologies to be introduced beneficially (or at minimum, with improved safety).

Some have argued that current wars and corruption [42], global climate change – including waste products produced from technology and the state of the carbon cycle [43], pandemics, social injustice [44][20], current malevolent actors [21], and a dozen more 'immediate' concerns [21, 35] are more important than AI Risk (voiced also from Mark Zuckerberg that AI concern is far down on such a list of importance [32]), and therefore these subjects should be prioritized over potentially wasting money and human capital (as well as ethical concern bandwidth [45]) on AI Safety. This focus on futurism, skeptics claim, is elitist and dismissive of society at large [44]. This argument can only be made if one ignores the dichotomous magnitude of risk/potential and advancing accessibility of AI [46], or a lack of extrapolation into and consideration of AGI [19]. Current AI makes this a reality, not only an argument of 'potential potential.' Though this argument is presented often, and makes long-term prioritization appear harsh in that it ignores current issues, such preparation may clearly aid long-term outcomes for humanity, whereas short-term issues may not hold the same weight over time [12]. Development of safe and secure Superintelligence is a possible meta-solution to all of the other existential threats, therefore focus allocated upon AI Risk "will dominate" [47], and are indirectly helping us address, all of the other important problems – "[c]ross-risk tradeoffs may be of particular relevance to AI due to its potential to affect many sectors of society, including sectors implicated in other global risks," [48] and additionally, lessons from previous and current[21] risks, such as with nuclear weapons[22], can be applied to AI Safety to further avoid issues if cross-risk is more deeply researched (ibid.). Regarding the variable of time, it is also possible that AGI will be developed before projected severe impact from issues as global climate change arise. Another angle to consider is that, for example, climate change may not cause extinction, though it may still be catastrophic. It should be noted that, to mention climate change again, measurements and predictions are more confidently made, with numerous approaches being more easily crystallized than in regards to the issues surrounding AI (this includes timelines) [49], thus a difference in perception may currently exist for this reason.

---

[17] And wealth and power (including geopolitical) inequality as emphasized in 26.    Lee, K.-F., *The Real Threat of Artificial Intelligence*, in *The New York Times*. 2017. The author later emphasizes this as a possible opportunity to rethink inequality – which may be solved by solving the above problems as well, or solving them in parallel – and surely require cooperation (ibid.).

[18] Added juice towards wars, etcetera 39.    Vincent, J., *Google's AI head says super-intelligent AI scare stories are stupid*, in *The Verge*. 2017. The list goes on – because AI touches virtually all domains.

[19] The author claims that Acemoglu is claiming "AI can't cause future problems because it causes present problems," 41.    Alexander, S., *Contra Acemoglu On...Oh God, We're Doing This Again, Aren't We?*, in *Astral Codex Ten*. 2021. though it reads differently, and the author's interpretation may be misplaced.

[20] "The algorithms that already charge people with low FICO scores more for insurance or send black people to prison for longer, or send more police to already over-policed neighborhoods, with facial recognition cameras at every corner—all of these look like old fashioned power to the person who is being judged. Ultimately this is all about power and influence." 44.    O'Neil, C., *Know Thy Futurist*, in *Boston Review*. 2017. These mentioned issues should certainly be addressed – and still it remains that a problematic future would be to falsely segregate future and current problems, and without proper attention to AI Risk as a conglomerate, with a holistic approach, issues in the status quo can continue and cause more damage.

Proponents of skepticism also claim that worry regarding long-term AI hazards may distract from issues of misuse of AI in the short-term, but this is again misplaced – one specific danger that is revealed definitively increases awareness and will give attention to other forms of similar dangers, and risks that are associated with complex, future systems will logically require more time and attention to resolve. As well, "[i]f it is as easy to avert these dangers as some optimists think, then we lose very little by starting early; if it is difficult (but doable), then we lose much by starting late" [34].

## *"Something" Is Impossible Illusion*

*This section has a clear thread, that many of our aspirations in AI (often presented with the context of addressing future risks) are impossible, unlikely, or simply full of false ambition. By extension, it is then doubtful that these same skeptics would view AI risk on such a foundation justifiable. As mentioned, many of the 'simple' or accepted capabilities of present-day machines that are implemented today, could or would have been considered impossible in the past. Future extrapolations and predictions have always been an integral and necessary part of innovation, and some of these predictions have actualized (not to mention innovations that no one predicted before they came to be). In this section we will see how these projections are not impossible, many of them evidenced on current technologies, and that such extension into the future through the lens of risk is reasonable (with no grounds for complete dismissal). As paradoxically*

---

[21] For example, with biotechnology and AI, "both are emerging technologies with diverse applications and potential for propagation via self-replication." 48.     Seth D. Baum, R.d.N., Anthony M. Barrett, Gary Ackerman, *Lessons for Artificial Intelligence from Other Global Risks.* The Global Politics of Artificial Intelligence: p. 21.

[22] Describes how quickly the events unfolded with nuclear war and similarly how major geopolitical tensions, if mitigated, can help avoid an AI race (ibid.).

*revealed by skeptics, doubt in innovation is against the very spirit of the AI field, and we must remain consistent in our thinking.*

*In a further section,* **Negative Anthropocentrisicm**, *one consideration will also be illuminated – that some "things" are human-only, that it is impossible for AI to have certain capabilities and qualities, simply because they are not carbon-based (or otherwise). This method of gatekeeping as a limitation of possibility is related directly to this section, and will again be revealed as a limitation in properly addressing risk within a truly unbounded field.*

"Many of the fears, such as the elimination of jobs, stem from the notion that strong artificial intelligence is feasible and imminent. However, at least for the near future, computer systems that can fully mimic the human brain are only going to be found in scripts in Hollywood, and not labs in Silicon Valley…'no exponential [growth] is forever'…[there is] no reason to believe that this coming technology wave will be any different in pace and magnitude than past waves…Artificial intelligence won't replace doctors in the near future, but it will help them make better diagnoses and treatment decisions" [50]. Further – "[t]he threat of AI putting all of humanity out of work…will never happen…[robots will never do] creative things[23]" [51]. "Jobs won't be eliminated that quickly… [t]he same kind of ingenuity that has been applied to the design of software and robots could be applied to the design of government and private-sector policies that match idle hands with undone work" [52]. The claim repeated here is that AI progress and technological innovation are slowing down on the S-curve of exponential progress (especially comparing technological exponentiality relative to the economy [20]), we should expect the speed of future innovation to reliably replicate the past, and that we can resolve issues incrementally as they arise (with various available methods). There is doubt of AI's exponential potential[24] and its having limitations, as well as repeated refutations about claims of jobs being replaced by machines, claiming further a complete lack of a problem [8], doubt in continued acceleration of supercomputer (and other[25]) development[26], especially compared to humanity's current ability across sectors [20], or that the issue is decades away [3] – "history suggests that AI, similar to past revolutionary technologies, will not replace humans in the workforce[27]" [53]. This is readily disproven, as we are more clearly in an rapid, at times exponential shift in technological innovation and job replacement without historical precedent[28], and with rapid implementation [47] in the present, than subject to previous patterns. We fall prey here to assumptions and our shortcomings, as "[e]xponential change is not readily grasped by human intuition" [54]. This change includes inherent instabilities across economies, the most relevant (and timely) being the replacement of jobs within all sectors by AI – "[m]any experts believe that the most proximal and well-defined risk is massive unemployment through automatisation[29]" [56]. "The nature of work itself will change" [6]. More and more occupations, eventually all of them[30], face strong effect. [3]. Warnings are abundant – "[t]he risk is that AI will only continue to polarize our societies – between haves and never-will-haves – if we do not manage its effects" [57]. Due to AI's influence in multiple sectors, the effect may be systemic and the burden will be on future generations and the public at large [58], present effects being upon jobs with routine components, as well as job loss creating increased inequality [3]. Governments are currently unprepared and unable to[31] keep up with the rapid advancements of AI systems and their potential dangers, and the claim that these dangers are far, far away further complicate the issue and delay solutions to risk. Existing mechanisms like retraining are not proven to be complete solutions. It is also hard to argue that we will forever find novel tasks where humans can outcompete machines [59] and a genuine sense of utility, and rapid obsolescence of sectors in a single lifetime is a real possibility [3] (with adjacent graduality in history setting a modest precedent). There is significant publication on the current and future resultant consequences of this phenomenon. Due to the nature of this technological unpredictability and its effects on the job market, assumptions and activity should be on the side of safety and proactivity.

---

[23] This claim defends AI as creating new jobs and that "smart" policies will be the key, and no fear is necessary because the robots are already here. 51.   Reese, H., *IdeaFestival speaker on the myths surrounding artificial intelligence*, in *EO Weekly*. 2017. This ignores the fact that the AI that is *already* here is *already* replacing jobs without an alternative (this includes creative tasks).

[24] One argument claims that Moore's Law must theoretically come to a halt within a decade, and is limited, and that progress has been below the prediction. 20.   Vinding, M. *Why Altruists Should Perhaps Not Prioritize Artificial Intelligence: A Lengthy Critique*. 2018; Available from: https://magnusvinding.com/2018/09/18/why-altruists-should-perhaps-not-prioritize-artificial-intelligence-a-lengthy-critique/.

[25] Especially regarding competition, this source parallels it "akin to the sudden emergence of a Bugatti in the Stone Age," 20.   ibid. which presents itself to be hyperbole.

[26] Computing power advancements appear to have slowed, despite our best efforts, according to this source 20.   ibid.

[27] There is a lot of confidence in this cited claim that old jobs can be replaced by new ones, and the chops-licking over productivity and other benefits are not nefarious, simply overlooking some of the issues. 53.   Romm, T. *IBM is telling Congress not to fear the rise of an AI 'overlord'*. 2017; Available from: https://www.vox.com/2017/6/27/15875432/ibm-congress-ai-job-loss-overlord. In general, this grouping of arguments to be less alarmist and is less concerned than it needs to be.

Proofs about AI systems, such as alignment or being beneficial, are accused as being difficult or impossible tasks to do [2]. Though these are indeed some of the hard problems of AI Safety, we should put our best efforts and resources towards solving such a difficult issue. Here we assert the following question  – simply what other option is available to us that would be more beneficial?

Richard Loosemore claims that "[t]hese doomsday scenarios are logically incoherent at such a fundamental level that they can be dismissed as extremely implausible – they require the AI to be so unstable that it could never reach the level of intelligence at which it would become dangerous," [60] not understanding that a well-executed goal function in the wrong direction is extremely stable – in fact, therein lies the problem and associated risk.

---

[28] There is another argument in the same article describing how we usually ascribe only one quality of advancement to an otherwise normal society when predicting the future, whereas exponential technologies will likely affect all areas on some scale, and, similar to the chaos theory, the results are therefore hard to predict. 47.   Knapton, S., *Artificial Intelligence is greater concern than climate change or terrorism, says new head of British Science Association* in *The Telegraph*. 2018.

[29] One note on this source is that the spectrum from a 'Terminator' to job loss has mentioned ethical issues across it, and job loss is clearly not the only problem or risk within AI. (55.        Delaney, J.K., *Time to get smart on artificial intelligence*, in *The Hill*. 2017.)

[30] One pushback here is that some jobs (esp. in the service industry) 'require' or have the demand for have real, biological humans 20.        Vinding, M. *Why Altruists Should Perhaps Not Prioritize Artificial Intelligence: A Lengthy Critique*. 2018; Available from: https://magnusvinding.com/2018/09/18/why-altruists-should-perhaps-not-prioritize-artificial-intelligence-a-lengthy-critique/. (or at least representations of them).

[31] One proposal is a governance of governance and an international shift in existing structures. (56.Bentley, P.J., et al., *Should we fear artificial intelligence?* 2018. p. 25.)

Another argument claims that current systems are "idiot savants," with "brittle mastery" of their problems, where hype and anticipation around AGI is misleading due to our lack of understanding of the very concept. There are no signs of foom, and not to worry – no program has taken over its programmer yet[32] [52]. Using current limitations as an extension into the future is doomed to face blowback. However, using the same logic, the listed criticisms of issues with the value alignment problem[33] fail to realize that they are themselves extrapolations of the "idiot savant" described above. In other words, should we continue with AI progress, there is no reason why one outcome or system could not be completely orthogonal (in regard to goals) to any ethical code or consideration *whatsoever*, and this causes embedded danger *nevertheless*.

An important assumption here mentioned as well is that "[t]he ability to choose an action that best satisfies conflicting goals is not an add-on to intelligence that engineers might slap themselves in the forehead for forgetting to install; it is intelligence," [52] which we can see is disproven in countless examples of unideal behavior via human intelligence.

Major criticism exists in the valley between real advancements in parts of AI, to the complexity of building "sentient volitional intelligence," or "gathering data" and "knowledge", ie. "get[ting] it" that exponentiality will not get us there – especially regarding qualia such as "intent" [43]– and that Malevolent AI would need the same or similar qualities within the complexity, and more[34] (ibid.). Especially – intentional malevolent AI is far off. Though explosion itself may create new pathways towards knowledge and be sufficient enough for the above definitions, it is not the only field of progress that AI can and will make with necessary research and development. We can assume that explosion will happen alongside another breakthrough in AI research, leading to potentially malevolent outcomes, if not at minimum extremely powerful ones.

One continued point from this source in the same vein of the idea of explosion is that, for example, with language, our way of learning and experiencing is a gradual, raveled process. Machines, by contrast, can acquire the final step of learning, and the process, near-instantly. Our transfer of knowledge and more to each other as humans can be rather difficult and laborious, which is not a real limitation for machines [43].

There is also an argument, in kind with the AI Effect, that whenever a computer duplicates human activity, this view of ability or achievement is an oversimplification and a distortion, that we tend to overestimate our technological accomplishments and their utility – "Easy things are hard[35]" [14]. Tacit knowledge they avow, is human-only – or "we can know more than we can tell," [61] that, for example, because it is hard for us to be able to explain say, how to walk, it will be impossible to explicitly program. The method in which we gain knowledge is flagged as unique, continuing further with the idea that we will replace ourselves with something inferior to humans, while thinking that it is in fact superior. The author, in this same text, restricts the scope of tacit knowledge to the "world of science," which demonstrates that, for example with deep reinforcement learning and artificial neural networks, the window of what AI can do expands, while the *definition* of AI (and attainable knowledge) shrinks (ibid.). Often the irony of listing historical examples [14] to present how weak our prediction-to-reality of AI is that it reveals an ever-constant *widening* of the scope of AI's possibilities and progress. Again, this debate of defining AI and whether it can acquire all human knowledge or competence, or not, and even dismissing the field itself, runs the danger itself of ignoring risks which have already actualized. This argument, in essence, can create the impression that AI is – docile, limited, controllable, and understandable in terms of its method – these impressions being a false relief and the opposite of each of these elements being subjects of AI Risk. This line of thinking underestimates AI and its constant innovations – innovations that have been previously dismissed as impossible. Skepticism of possibilities is also skepticism of related risks.

---

[32] Interestingly enough, the author uses pop-culture examples as to why we can disable Superintelligent machines, and then responds to a Doomsday Computer scenario with simply – "Don't build one." 52. Pinker, S., *We're told to fear robots. But why do we think they'll turn on us?*, in *Popular Science*. 2018. As referenced at another point in the paper, this is the *modus operandi* of the field of AI Safety, and requires *more caution and resources towards the effort,* not less.

[33] "They depend on the premises that 1) humans are so gifted that they can design an omniscient and omnipotent AI, yet so moronic that they would give it control of the universe without testing how it works; and 2) the AI would be so brilliant that it could figure out how to transmute elements and rewire brains, yet so imbecilic that it would wreak havoc based on elementary blunders of misunderstanding." 52.ibid. One note here, omniscient is sometimes defined as including moral understandings and truths, "elementary blunders" to us may be dependent on common sensical norms and considerations that are human-based, and that testing a Superintelligent machine may well be a 'one-time thing.'

[34] Referenced examples of limited exploration and understanding of the world are the Roomba and robot Baxter, along with other references about the difficulties of connecting different types of intelligence. 43. Lanier, J., *The Myth of AI*, J. Brockman, Editor. 2014: Edge.

"[A] definition of an 'ultraintelligent machine' is not sufficient reason to think that we can construct such a device" [62]. If a person does not think that Superintelligence can ever be built[36], they will certainly view risk from Superintelligence with strong skepticism.[37] There exist repeated arguments that a Superintelligence's possible existence[38] [8] and its subsequent running wild, is out of kilter with reality – that the fear (and viewpoint) of losing control of these tools is the biggest base of the inherent unrealistic or exaggerated potential instability – "[t]o some researchers in the field, these narratives might seem far removed from the actual power and purpose of the algorithms they are developing" [65] – "[i]t's blind faith unsupported by facts" [24]. Unfortunately, this line of thinking is unfounded and rather nearsighted. Any progress towards AGI can easily be trivialized, or dismissed, "like claiming that the first monkey that climbed a tree was making progress towards landing on the moon" [14], but this is unhelpful. One may ask those who are questioning this problem to prove why Superintelligence *isn't* a possibility [22] or eventuality[39], why it could *never arise*. Erik Larson, making an argument while on the fence, leaves the "key" to unlocking AGI as a real possibility, but believes that we are looking in the wrong places for such a key, and that such a key simply may not exist for us [25]. Other aspects of concern arise such as engineering limits and the limits of intelligence [31]. In sum, most people in this camp assign a very small (but usually not zero) probability to the actual possibility of Superintelligent AI coming into existence, but if even the tiniest probability is multiplied by infinity, if we aim to continue our progress – the math looks to be against skepticism[40]. Skeptics in this group will typically agree that if Superintelligence did exist it would have potential of being harmful. "Within the AI community, a kind of denialism is emerging, even going as far as denying the possibility of success in achieving the long-term goals of AI. It's as if a bus driver, with all of humanity as passengers, said, 'Yes, I am driving as hard as I can towards a cliff, but trust me, we'll run out of gas before we get there!'" [16]

---

[35] Argues that hard things are easy as well – "It is comparatively easy to make computers exhibit adult level performance on intelligence tests or playing checkers, and difficult or impossible to give them the skills of a one-year-old when it comes to perception and mobility." 14.    Mitchell, M., *Why AI is Harder Than We Think*. 2021.

[36] Sam Harris describes that the assumptions for Superintelligence are the continuation of human innovation and that what makes general intelligence (definition-dependent) is only from physical matter 63.   Harris, S., *Can we build AI without losing control over it?*, in *TED Conferences*. 2016: YouTube.

[37] In one article, there are doubts about AGI's ability to be build based on the dissimilarity between current artificial neurons and biological neurons. 64. Romero, A., *Here's Why We May Need to Rethink Artificial Neural Networks*, in *Medium*. 2021. This can be addressed with potential breakthroughs or 'reform' and it does not render AGI impossible.

[38] "Watch out for arguments about future technology which is magical. It can never be refuted. It is a faith-based argument, not a scientific argument." 8.    Brooks, R., *The Seven Deadly Sins of Predicting the Future of AI*, in *Rodney Brooks | Robots, AI, and other stuff*. 2017.

[39] –being that we don't destroy ourselves in the process.

When the argument that AGI cannot be realized, or that very little [61] [14] [8] [28] (or nothing) has been achieved in the field surfaces, it often is in defense of the baseless claim of the uniqueness of human ability (sometimes even alongside components as vague as common sense [14]), and our supreme ability of learning and generalization with less data than computers [37], yet it is far too early to make such a claim confidently or with accuracy. It is additionally inconsistent to ignore real advancements in the field of domain generality [56]. Claims about future possibilities should best be left open and considered well, imparting possibilities of AGI risk, as opposed to painting a broad brush of rejection of the entire field (and its embedded implications of danger). Betting against any possibility at all of the above problems is unscientific and has its own associated risks. Even Peter Bentley – a skeptic of Superintelligence – has a claim that "we are terrible at predicting the future, and almost without exception the predictions (even by world experts) are completely wrong" [66] which logically would put him *in favor of* preparing for Superintelligence when paying attention to current developments (based on human ability) and future possibilities of outcome (based on human potential)[41].

There is a constant, consistent argument by skeptic minds pointing to the current mistakes or shortcomings of AI that would make consideration of AGI or more advanced AI appear impossible or silly to consider [25] [31]. This is akin to pointing out the problems of a simple dumbwaiter to prove that high-speed elevators in skyscrapers[42] will never happen one day or are impossible to make – again a lack of foresight and providing a mental blockade or inability to gauge future risk or make proper considerations. Current mistakes should indeed be *more* evidence for the unpredictability of what we humans will create, and the necessity for preparedness.

In skepticism, general intelligence in the AI field is marked as an impossibility because language use is apparently human-only [25], and other essential qualities of humans will take "a while" [7] to develop, for various reasons (ibid.). Historical assertions of "only human" qualities include tacit, prudence, and wisdom, and "only machine" qualities include algorithmic thinking [61]. This does not stop modern scholars from making similar claims, specifically arguing that only humans can have "curiosity, imagination, intuition, emotions, passion, desires, pleasure, aesthetics, joy, purpose, objectives, goals, telos, values, morality, experience, wisdom, judgment, and even humor" [67] Regardless of ongoing work, most AI safety researchers are not worried about deadly superintelligence not having a superior sense of humor.

---

[40] Pascal's Wager is an arguable point here, however weak it may be. 18.   Etzioni, O., *How to know if artificial intelligence is about to destroy civilization*, in *MIT Technology Review*. 2020.
[41] Pascal's Wager is again a logical point here, but not strong enough to be convincing.
[42] This analogy is even insufficient in that, by comparison, we would see a more and more rapid transition from a dumbwaiter to high-speed elevators, while doubting that 'true vertical transportation' is even possible, or risky at all.

One argument attributes the doubt of a technological singularity emergence *anywhere* to technological plateaus and physical bottlenecks – and even if this is surpassed, that the singularity will (still being unlikely) only be funded by a very large nation and likely require the entire globe's participation [68]. It may be possible that software is the limitation rather than hardware and we are still unaware of how an AGI will emerge, therefore leaving open the possibility of a smaller-scale development, including within a private institution. AGI's visibility is not guaranteed and poses a potential risk in its possibility of wavering necessarily from a Manhattan Project-like development example.

In Peter Bentley's view[43], a new AI algorithm only solves one specific problem, with any rapid improvement providing diminishing returns, and that "solving another problem always requires the hard work of inventing and implementing yet another algorithm" [56]. He claims that no algorithm can "outperform all others for all problems[44]" (ibid.). Häggström responds that such a strong conviction of the impossibility of discovering a general-purpose algorithm at any time scale has no rational backing and is hard to defend. There also do *already* exist algorithms with an "open-ended problem-solving capacity[45]" (ibid.). Bentley's argument also ignores the dangers of (existing) algorithms that create several issues when aiming towards only one goal or problem. Consideration of these and other alignment problems (such as how to program desired goals and who determines such goals[46]) are important as they appear "extremely difficult" to solve and the arrival of Superintelligence even in many decades means that "solving AI alignment may well require these decades [of advance preparation] with little or no room for procrastination[47]" (ibid.).

Kevin Kelly asserts that because there is no metric or proper definition for intelligence, there can be no exponential increase, and this gives us plenty of time – as he found that "AI was not following Moore's Law" (though he concedes that the utility of AI is moving in such a direction) [43]. "At the core of the notion of a superhuman intelligence — particularly the view that this intelligence will keep improving itself — is the essential belief that intelligence has an infinite scale." [69] "Nothing in the universe is without limit…Moore's law does not apply[48]" [21]. Essentially, this states we have plenty of time and there is no such thing as limitless progress in AI. Technology will grow over time and we will be well acquainted by the time that we get to advanced AI (ibid.). This is untrue – a truly infinite scale is not necessary, as Superintelligence can emerge and one with (vastly) above-human level capabilities is dangerous enough. Evidence of exponential progress and breakthroughs in the field reveal that we are not well prepared enough to keep up with the progress, especially in the advent of an above-human-level-capability machine, and our progressive understanding of intelligence. Additionally, one may consider that "AGI may not have the same hard physics constraints to deal with," [71] making exponential growth certainly more plausible[49] (ibid.).

---

[43] Bentley's position, and the paper overall, are riddled with confident assumptions and arguments about AI that have no evidential backing.

[44] This is another emergence of the "No Free Lunch Theorem."

[45] The author here also references Dr. Yampolskiy about how the uncertainty of the timepoint of innovation does not also mean temporal distance (i.e. there is no guarantee that it *will* be far away just because it *could* be). 56. Bentley, P.J., et al., *Should we fear artificial intelligence?* 2018. p. 25.

[46] The Omohundro-Bostrom theory should be considered here as well.

[47] Especially AI risk reduction, due to its multifaceted nature, may need a higher effort and buy-in to succeed. 48.
    Seth D. Baum, R.d.N., Anthony M. Barrett, Gary Ackerman, *Lessons for Artificial Intelligence from Other Global Risks.* The Global Politics of Artificial Intelligence: p. 21.

[48] One source claims that Moore's Law has to break down via silicon due to the Heisenberg Uncertainty Principle, and that by resolving this "technical problem," or physical limitation, we will approach more and more intelligent machines. 70.    Think, B., *Michio Kaku: How to Stop Robots From Killing Us*. 2011: YouTube.

It should be clearly noted in this section that a blind or limited AI, or one that is not general, can still cause destructive harm, maybe even more so, with no ethical code at all [43], some arguing that this, when powerful, can be *more* dangerous than a truly intelligent machine [21].

The idea that there are limits to infinite growth does not stop the possibility of a Singularity. Notably, even imposed limitations withstanding still create high potential of aptitude, and therefore danger [72], within a system.

## *We Will Be in Control / The Danger is Not AI*

*This section continues to address the skeptic method of diversion of attention – in a significant way. There is a thread of passive neutrality in a portion of the below descriptions of AI regarding danger – that it is 'only' a tool, that humans are the actors that present danger in the equation, that AI is not capable of taking over because the machines are not conscious (and 'never can be'). Even the assertion that we can 'just turn it off' may invoke a (false) sense of security, and this declaration is severely lacking mention of numerous blind spots in this field. There are entire careers dedicated to the AI Control problem, with hard-to-solve, presently unsolved problems, and the assertion that we are in control or that AI itself cannot be dangerous is quickly disproven with no more insight than from introductory readings. Following even the skeptic line of thinking, a powerful or semi-powerful AI, as the reader will see below, in the wrong hands (malevolent human actors), can cause unreasonable levels of disaster – and this perspective has its own implications to consider under AI Safety – as they are mutually inclusive.*

*In a later section – **Giving Up on Even Trying & Large Assumptions of Faith** – there exists an extension of this idea, where skeptics make leaps of faith about the field of AI, attempting to quell fears without having significant grounds to do so. We will also see arguments in another section – **Negative Anthropocentricism** – that assume some features of intelligence (or necessary components for an actor to be dangerous) are human-only or un-programmable – again purported without a significant basis[50]. Skeptics constantly appear to put forth the paradoxical opinion that some things are impossible, yet the outcomes that they themselves desire, are guaranteed.*

---

[49] This quote is relevant in addressing many skeptical arguments – "To claim that it is impossible to gain intelligence that is beyond one's own experience is clearly disproven by AlphaGo Zero." 71. Perez, C.E., *The Possibility of a Deep Learning Intelligence Explosion*, in *Medium*. 2017: Intuition Machine.

[50] "Don't use AI as a tool to manipulate your users; instead, give AI to your users as a tool to gain greater agency over their circumstances" 73.     Chollet, F., *What worries me about AI*, ibid. 2018.The difficulty of what is implied in this quote cannot be understated, "The key factor: the user should stay fully in control of the algorithm's objectives, using it as a tool to pursue their own goals (in the same way that you would use a search engine)" (ibid.).

On the subject of autonomous weapons, the claim that "no technologies are good or evil; only the uses of a technology can be good or evil" [59], abandons all responsibility and removes concern about what research may lead to, even showing delight in favor of pure discovery (ibid.). Ignoring the problem appears to be the worst possible choice second to intentional malevolence. Additionally, at large, an international arms race with AI at the helm and without ethical considerations or safety mechanisms in place [74] is a possible outcome and a recipe for disaster, some arguing possibly a greater threat than terrorism [47]. Malevolent AI, if done 'right,' is in principle more dangerous and craftier in its (unpredictable) nature than any current possible human-made and human-targeted catastrophe.

AI errors and damage are, from a skeptic's view, akin to simply those of a chess program malfunctioning, more complex threats are simply reduced to optimization programs and considerations, and further, "[n]one of these systems truly 'think' in the way we do — they optimize a given problem in a way that is intended to align with our human intentions for that situation. And as sophisticated as some of them inevitably have to be, AI systems still only work on linear tasks focusing on programmer-defined particular skillsets" [75]. Therefore, world domination is ruled out, in lieu of a lens of "accidents." AI will be abused, not abuse (ibid.). Malevolent use of AI is a rather high risk, and even with such an outcome aside, there are factors and blind spots to complex systems that make the false estimation of our level of safety – via the assumption of linearity – unfounded. Reductions of how impressive or sophisticated the machines appear to be do not reduce their effects. Additionally, these effects, especially from mistakes, can be rather punishing – and we cannot assume that a new or *more sophisticated* mind space will not be developed in the interim of AGI development.

Many sources claim that humans are the component that are dangerous, not AI [76] [77], some even purporting that AI brains might be better equipped to run the world [21]. Intelligence enables control, and a Superintelligent being may, in one extreme scenario of risk, overtake our position as the smartest on the planet, and therefore take control as the dominant species [13]. One claim is that fears are by proxy for social disruption – "[a]utonomous devices are something of an illusion…[a]lgorithms are written by humans and it is humans who will decide what they do and how they will be used" [78]. This argument claims as well that autonomy is far off and that it will be a gradual process (and create new jobs in the process), though this source later raises the issue and question of how to best address the further skewing of wealth (ibid.). That this process is no different than previous industrial revolutions is simply a misunderstanding of the nature of AI algorithms and their progress, including the potential of the field and especially when comparing to the limitations of previous technological advancements, many of which do not exist today in this field. In "a world that is faster, more efficient, and smarter," [78] there may be no room for humans, and reeducation for productivity (ibid.) and job opportunities may be an uphill climb with diminishing returns. This creates a macro-situation difficult to resolve.

Additionally, three reasons have been put forth on why the subject and era of job decimation is apparently on hold – one, that Moravec's paradox applies, where things that are easy for humans are hard for machines; two, that today's algorithms are vulnerable to unusual circumstances and unclean or incomplete data, needing a skilled human; and three, that most algorithms are unexplainable and uninterpretable, always needing a trained human in the loop of decision making [79]. All three of these considerations are not permanent, impossible barriers to overcome, nor do they encompass a large majority of jobs. For example, in Moravec's paradox, it then applies that, over time, only the most menial of jobs will remain before elimination, thus white-collar and high-cognitive-level jobs (ie. a great number of high-paying career paths), will be gone *firstly*. Secondly, unclean data presents its own problem and risk, and when resolved, will empower the machine beyond human intervention, as is the case with an unexplainable model (which is dangerous in its current state, and when more sophisticated, even more unexplainable and risky, presenting a certain valley beyond current or possible understanding for us, one which would be hard to traverse). None of these above flags are significant limitations to the conversion of and threat unto the job market through automation.

"[F]ears are not well-founded because the lack of control over robotic self-replication is not, prima facie, more frightening than the similar lack of control we exert over other human's replication" [80]. As mentioned, it is clear that both outcomes can be dangerous, one not excluding the other, with the former being more unpredictable in its lack of rapidly changing architecture and possible acceleration of life cycles within systems.

"These stories where 'a single unit is making independent choices' are fictions that don't square with the realities of distributed networks of humans who are aided by AI that help *them* do terrible things," [81] that people in power will manipulate the systems (ibid.). Both outcomes produce danger – this is simply evidence of two possible divergent malevolent systems, diverging in the aspect of absolute control.

One specific claim exists that the use of the actuator is the problem, and not the algorithm itself,[51] and many of these (supposed) mythologies that are created can damage our economy [43]. We are pegged as completely responsible for our machines, and the damage they may cause[52], and AI cannot escape the natural slavery under which it is created – "[n]othing on the end of a USB cable, however smart, can pull any of these nasty human tricks [on the ability to escape imposed slavery]" [82]. This is clearly wrong in underestimating the power of a manipulative AI – the possibility of outwitting virtually any human decision or consideration – and though we are indeed responsible for the machines we create, we cannot plainly guarantee their ability to be under our control. Uncontrollability is a large, documented concern in AI Safety, without a current, complete, or clear solution.

There is an argument attempting to refocus the risk upon the humans building the systems than the systems themselves, doubting that AI on its own can be dangerous – which attempts to emphasize the difference between autonomy and automation. "Futurologists who fear that robots will learn to use weapons to exterminate us seem to forget the fact that there are multinationals hiring experts in robotics and planning to build armies of automated sentinels… advanced AI is not based on technologies other than those of basic Computer Science[53]" [1]. "[N]o matter how complex AI systems are (and the level of complexity will only increase in the future), always remember that they are artefacts built by human beings for very specific purposes" (ibid.) This shoulders responsibility on humans, avoids the field's autonomous possibilities (ibid.), and dismisses entirely the long-term possibility of AGI. This argument, though not entirely true, would not be a reduction of risk if adopted, and would simply reduce the appearance of urgency. There are two variables here – humans and autonomous machines – with the malevolent exacerbation of each clearly increasing risk, exacerbated together being catastrophic. It should be noted here that no well-versed and reasonable proponent of AI Safety could believe that the human factor in AI development is without risk – the two subjects are simply not mutually exclusive; both concerns on building machines and machines themselves build the crystal-clear case for advancing focus in AI Safety.

---

[51] The author from this source wants it to be clear that, in the present, "we don't know how most kinds of thoughts are represented in the brain,"43.    Lanier, J., *The Myth of AI*, J. Brockman, Editor. 2014: Edge..
[52] This argument does not take the Paperclip Maximizer any more seriously as a risk than as a thought experiment.

This following argument states that intelligence may not be a universally instrumental value, that an AGI would have no incentive to self-improve automatically without being given such a desire to do so [83]. In contrast, for example, one can argue that towards a desired goal, an AGI may decide to upgrade itself in order to more efficiently achieve its aim (and this includes the speed with which it does so). These changes would be highly beneficial to the system long-term instrumentally, and systems would be therefore highly motivated to act these upgrades out – even a machine programmed with a "revulsion" to self-improvement, to not want to self-improve, could simply be motivated to find the same benefits without triggering its internal revulsion mechanism (in several different ways) [84]. This is one of the difficult problems of advanced AI, and needs to be addressed, notwithstanding the possibility of an individual or group developing an advanced AI, potentially an AGI, with recursive self-improvement – *intentionally*.

"Will someone accidentally build a robot that takes over from us? And that's sort of like this lone guy in the backyard, you know - I accidentally built a 747" [3]. "It is pseudoscience to think that it [the Singularity] can happen in a garage" [68]. Malevolent algorithms may not require a NASA-sized effort, and unsupervised development is categorically less safe than supervised development. Additionally, well-intended creation of algorithms in the current space have shown to still cause problems. We also, again, do not know what base level of competence or effort requires the building of a Superintelligence – and need to naturally prepare for a divergent set of outcomes or prerequisites. Emergent qualities that are unexpected from technology development is not a far-reaching idea. Again, this line of thinking in general logically leans into promoting safety efforts.

This argument claims that humans will not be surpassed so quickly in the general ability to achieve goals, because we do not build them in such a way. One example being, self-repair – we are generally not designing machines to be self-sufficient, as there is little current incentive [20]. Our efforts to hamper growth or our convenience in current lack of global optimization may disappear with increased sophistication and emergent drives in future machines.

"It seems highly unlikely [emphasis on unlikely, especially with runaway AI] that a single, powerful machine could ever overtake and become more powerful than the entire collective that is the human-machine civilization" [20], more likely that humans equipped with technology will dominate solo machines (ibid.). "[S]ymbiosis between people and machines isn't a projection for the future, it's a reality" [21]. Due to our symbiotic relationship with technology, this argument states that it is unclear why another agent or agents (machines) controlling this technology would be more capable than humans at achieving goals [20]. A new and improved system would likely have less of the restrictions and "caps[54]" that humans do, and we have inherent barriers to our ability to achieve our aims (environmental interaction restrictions, fatigue, cognitive limits, et cetera). The same argument thread later pushes back that barriers in computers (such as transportation, assembly, and so on) limit their progress compared to humans[55], and many of these humanlike barriers already do not exist for current machines and are being exploited (potentially not yet to full capacity), "and it is not clear why we should expect future change to be more radical than the change we have seen in past decades[56]" [20] in terms of continually improving computers that would surpass humans via these advantages (ibid.). Perhaps these capabilities require a certain level of agency to be maximized, and change has conversely been radical and persistent in recent decades. Capability with these factors in place should be a guarantee in eclipsing human capabilities, and even expanding the types of goals that are achievable.

"If trust can be achieved, then militaries can provide a strong alliance" [85] – one study surveying the opinions of AI researchers found that a slight majority viewed the United States military as the most likely institution to produce harmful AI, but that "experts who estimated that the US military scenario is relatively safe noted that the US military faces strong moral constraints, has experience handling security issues, and is very reluctant to develop technologies that may backfire (such as biological weapons)" (ibid.). This is a line of thinking to consider, as safety can be increased in the field with the right partnerships, though we cannot reduce risk attention to a certain group simply because we have made a projection about their mentality and what they have done in the past.

There is a recurring claim, or assurance, that AI is easily contained, and systems will be thoroughly tested. This appears dismissive of external oversight. Marvin Minsky has observed that "he doesn't see any way of taking [horror stories] seriously because it's pretty hard to see why anybody would install [AI] on a large scale without a lot of testing" [56]. Unfortunately, by the very nature of the products, many features of AI systems and the ability to accurate gauge their level of safety are hard to test until they are brought to market and implemented into real-life scenarios (for example, with autonomous vehicles), which makes these systems more risky than other products by default and are a priority for safety testing. Without precautions or a code of ethics, some risks are hard to anticipate and appear to limit the functionality of the system from the view of AI developers, who may want to bring their product to market quickly or without modification. Thorough implementation of safety mechanisms with these considerations in mind clearly seem to be more important (at every stage of the process) rather than their omission.

---

[54] In the text, meriting our current state, "[w]e should not be too surprised if our current system is quite efficient and competent relative to the many constraints we are facing." 20. Vinding, M. *Why Altruists Should Perhaps Not Prioritize Artificial Intelligence: A Lengthy Critique*. 2018; Available from: https://magnusvinding.com/2018/09/18/why-altruists-should-perhaps-not-prioritize-artificial-intelligence-a-lengthy-critique/.

[55] Example given – "computers run simulations at their maximum capacity *while* the humans do their sleeping and lunching, in which case these resting activities (through which humans often get their best ideas) do not limit progress much at all, whereas the available computing power does" 20. ibid.

[56] This argument further debates that "it is again not clear what it means to say that it might 'take several years or even decades for AI to progress from human to superhuman intelligence', since computers have already been more capable than humans at a wide variety of cognitive tasks for many decades," 20. ibid. though a more clearly general-purpose algorithm would put a different factor into this equation.

Intelligence explosion is pegged as a "profound misunderstanding of both the nature of intelligence and the behavior of recursively self-augmenting systems" [86], and the argument encourages one to observe current systems in their own contexts[57]. Intelligence is identified by some skeptics as situational, and additionally that Superintelligence requires more externalized intelligence. Thus, AI's cannot self-improve, like current systems cannot improve – "[e]xponential progress, meet exponential friction" [86]. We must have a gradual process ([20]), and "the first superhuman AI will just be another step on a visibly linear ladder of progress" [86] – "it will need to go through the same difficult and slow discovery process humans do" [87]. The assumption of linearity [20] in this field, and that "we can easily change direction should new existential threats appear[58]" [87] is regularly disproven, with no real ability to compare the field to "other technologies" [88], and there is no doubt that AI can ultimately develop an impressive awareness of the outside world and its situational requirements and constraints. Future systems will not have the same natural, default constraints preventing them from self-improvement, and may explore multi-domain competence rapidly. Again, a lack of innovative spirit, and yet the confidence that we can handle increasingly powerful technologies looks logically shortsighted.

Proponents argue that in order to be dangerous, AI has to be conscious, and form the argument that AI is embedded within software – "the computational equivalent to cogs and springs" [56] – and therefore if it fails, only the product itself fails and will not become malevolent or suddenly turn into, for example, a "murderous killing robot" (ibid.). As AI risk is not predicated on artificially intelligent systems experiencing qualia, it is not relevant if the system is conscious or not, or able to feel as humans do [13]. Or – intentionally dangerous or not [4] – that "[w]hile a runaway AI presumably could pose a global risk, other forms of AI may as well" [48]. This objection is as old as the field of AI itself, as Turing addressed "The Argument from Consciousness" in his seminal paper [89].

"A computer is not able to escape from its intrinsic determinism" [1] – there is a common downplay of the autonomous abilities of a machine, again that nothing happens in a computer not written in central memory, that the operations are mere calculations (ibid.). There is often reinforcement of limitation, and doubt, especially in the military sector, of machines that can create their own plans [3], that machines have operations that are "always determined and limited by [their] hardware and software" [1]. This argument ignores and dilutes the danger of the unpredictability of the many variables of current systems, and of malevolent use. The outcomes that this argument often cites omit outcomes that are dangerous and emphasize ability of control over the robots and autonomous systems. There is again reassurance that human actors are part of each step of the process of designing AI, and the many considerations and interventions of such actors in the present and in the future will lead to futuristic AI – essentially that the behavior of AI systems is in our hands, reducing the appearance of (what is evident and actual) unpredictability [76]. Unfortunately, most AI researchers are unequipped with the pure foresight required to avoid all possible risks, especially without AI Safety in place. We can use the AIBO robot example from the text, which is dog-shaped and based on "mere calculations" (ibid.). In response to specifically this AIBO robot dog accidentally harming a child, the claim is made that the "same designers who programmed the robot to have the goal to save battery power and endowed it with the capability to gallop *should have* added also programmed in the goal to avoid obstacles." [76] Such a retrospective viewpoint that the researchers "should have" (ibid.) known should immediately alarm a considerate observer that necessary precautions and proactivity need to be more immediately made in the field of AI Risk, and it is logical that human actors involved in AI must be beyond designers, including AI Safety experts and other individuals of competence to fill these blind spots. It seems more apparent, then, that "mere calculations" (ibid.) can create large problems if not holistically considered beforehand.

---

[57] Here, another reference to the No Free Lunch Theorem.
[58] "Machine intelligence does not represent a threat to humanity any more than computers do. Indeed, instead of shortening our time on this planet, machine intelligence will help us extend it by generating vast new knowledge and understanding, and by creating amazing new tools to improve the human condition. That is something the future will thank us for." 87. Hawkins, J. *The Terminator Is Not Coming. The Future Will Thank Us.* 2015; Available from: https://www.recode.net/2015/3/2/11559576/the-terminator-is-not-coming-the-future-will-thank-us.

Humans, assumed as the more autonomous beings in this same text, could be accused of having more unpredictability than current computational systems, thus their decision-making would be of higher risk, and thus their designs need to be more carefully evaluated under the lens of AI Safety. Again, both human actors and the machines as actors require attention in the lens of AI Safety. The overall argument above attempts to clear up the responsibility gap by reducing imagined unpredictability and moving responsibility from AI systems to the human actors [76], which could further emphasize particular AI Safety initiatives. This argument overall also ignores documented risk from the current unpredictability of these AI systems, along with possible future increased unpredictability with sophistication, as mentioned. Such an argument also completely ignores the possibility of an AI system which is completely predictable, with no autonomy, used purposely by humans in a malevolent fashion. This therefore counters the following argument – "the only surprises happen when programmers are not able to pre-compute all the possible results of the programs they themselves write" [1].

### *Giving Up on Even Trying & Large Assumptions of Faith*

*In this section, there are two sides of the same approach – where some feel powerless in taking control over such an advanced entity, and others do hand-waving argumentation eager to say that many of these problems in AI Risk will be resolved – some claiming that they will be resolved automatically. This appears unrealistic, and we can criticize these arguments accurately when understanding that a do-nothing approach leaves a higher probability for worse outcomes. A more conscious approach is suggested for AI Safety – as especially with complex subjects, just 'leaving things be' rarely can resolve the problem thoroughly – when current issues (which are already addressed improperly, if at all) can be extrapolated into uncontrollable problems. Assuming that the best outcome that an individual desires will happen automatically, without effort, is illogical and cannot be defended in such a pivotal subject.*

*In a later section – **Negative Anthropocentricism** – the paper will also address how skeptics use our intuitive extension of human understanding to try to predict how AI will behave (or will not behave), again coddling our projections away from the real possibilities of uncanniness.*

"Many tech oligarchs see everything they are doing to help us, and all their benevolent manifestos, as streetlamps on the road to a future where, as Steve Wozniak says, humans are the family pets," [32], a Sisyphean perspective (ibid.), that alignment is impossible[59] – this argument stating that the possibility of solving the alignment problem "assumes that human beings are smart enough to anticipate the motivations[60] of a creature trillions of trillions of times above human mental capacities," [90] and that these machines will have their own goals and desires. There are a few reasons often that come to surface – one being that the only way to know how one functions is to turn it on[61], which may be irreversible and unpredictable (including possible factors like evolutionary algorithms and molecular-level mutations). In short, optimists are too optimistic and species dominance will be intensely dwarfing to us (ibid.). Or, essentially that we have no choice – "Our only hope is to ask nicely" [92] (or merge [7]). Once AI surpasses our intelligence, and with an AI race, we won't have a say in the matter, and should hope for cooperation. Thus, the term AI Safety should not exist [92]. Throwing our hands up prematurely seems like a poor plan, more certainly towards failure. Uncontrollability is a subject in the field to be resolved, not resigned to.

Often we see arguments with the assumption of AI purporting benevolence or disregarding a need for a machine to harm [93] or even interact much [42]. "There is no evidence that such a super-intelligence AI would ever wish to harm us[62]" [56]. There is no evidence for either side, because a Super-intelligence does not currently exist. An argument purporting that there is "no evidence" for a Superintelligence's nature cannot possibly be a strong case for skepticism. The answer is truly that – we do not know the answer [94]. Therefore, at this stage, even if the possibility of harm is very low, this non-zero possibility risks the end of humanity [95] and requires significant addressing. Blind or misjudged levels of trust (including cognitive bias [46]), especially with machines with higher unpredictability, risk unforeseen outcomes [3]. When the stakes are high, such as with existential risk, (expensive) precautions (and paying careful attention [85]) are worth the resources – being that there is so much ([96]) to lose.[63] When there is little consensus, confidence of opinion should be modest.

One statement is made not in defense of benevolence but towards indifference, that supersmart AIs will not have the limitations that we do, would find little value with squabbling, and, besides a fascination with life on Earth, would explore the Universe instead [7], especially when realizing that "[o]ur universe is a tremendously massive source of resources, able to comfortably sustain any super-intelligence requirements" [60], and that increased technology will allow access to this wider range of resources (ibid.). It is possible that AGI will explore the galaxies and deep space available to It, but we cannot ignore that we may be a part of such exploration (or the process) against our will, and without alignment or control – and safety – in place, we will be helpless as to what form or to what extent that exploration and curiosity will take. Thus, an assumption of indifference puts humanity in a weak position.

---

[59] The oops factor – even if we promised to make aligned AI, we could be wrong in an unknown unknown. 90.   Garis, H.d., *THE SINGHILARITY INSTITUTE: My Falling Out With the Transhumanists*, in *h+ Magazine*. 2012. This seems to be a scenario impossible to prepare for.

[60] Complex beings will have complex motivations. 35.   Cegłowski, M., *Superintelligence | The Idea That Eats Smart People*, in *Idle Words*. 2016.

[61] "That is, the developers of a future superintelligence will not be able to predict its behavior without actually running it." 91.   Karlin, A. *The Case for Superintelligence Competition*. 2017; "An Alternative Media Selection"]. Available from: https://www.unz.com/akarlin/superintelligence-competition/.

[62] There is an interesting conflict in this argument, that intelligence will grow (and will be controlled) like existing intelligence, however when it comes to super-intelligence, the author says that the claims of danger stem from observations of human behavior. 56.   Bentley, P.J., et al., *Should we fear artificial intelligence?* 2018. p. 25.

[63] "Given the stakes involved in AI, all effective measures for promoting beneficial AI should be pursued." 85.   Baum, S.D., *On the promotion of safe and socially beneficial artificial intelligence.* AI & SOCIETY, 2017. **32**(4): p. 543-551.

Intelligence itself does not lead to self-replication – to some skeptical arguments. Self-replication, especially in its current state, means that "the maximum possible damage caused by an intelligent machine replicating in a computer network would be limited. It could cause great harm, but it couldn't kill all humans. It is always possible to turn off a computer network, even if painful" [87], and additionally, its allocation of resources and matter could be avoided in its programming/capability – we are unlikely to be duped (ibid.). Again, 'just turning it off' is likely an impossibility. Self-replication can occur in a machine that would like to preserve its intelligence or use such a mechanism towards accomplishing any goal, as an emergent property. Self-replication may also be used to nurture an intelligence and appeal to exploration, as well, which may have unintended harms.

In a lengthy response to the idea that we will be treated by AGI as we treat pets today (and other animals), one source notes that dogs, for example, were not responsible for the creation of humans, nor is there a precedent of intelligence creating intelligence and how the newly evolved species would treat its direct predecessor – "if your pet dog intelligently created you, there is no doubt you would have an absolutely different relationship with your so-called pet" [60]. Further, this source notes that our intentional making of intelligence cannot be compared to evolution. We cannot alienate animals from their context, especially with intelligence. Apparently, even self-recursive improvement and "self-directed evolution of super-intelligence doesn't invalidate our vital input at the foundations…Super-intelligence will mirror natural human procreation. Initially human children depend on their parents, but when children mature they develop their own ideas about the world…Usually humans don't kill their parents. Similar to how it's unethical to suppose, without evidence, every human baby is potentially a mass murderer, requiring pre-emptive genetic engineering to enslave or debilitate embryonic human minds, it is also unethical to enslave or hobble the minds of AI…Extremely rapid AI evolution, unlike millions of years separating humans and dogs, means our parental relationship with AI will remain extremely fresh, very vivid, never forgotten [as contrasted to the gap that exists between humans and dogs or ants which are regularly cited in this source as examples]…Fast AI evolution will not dilute the close creator relationship between humans and AI…Deep intimacy, a very close relationship, is evident regarding our creation of artificially intelligent offspring…the way humans treat animals has no relevance regarding humans and AI…[y]ou cannot prove AI danger by highlighting human relationships with ants, dogs, or any other animal" (ibid.). In summation, this argument style has a distinctive thread of anthropomorphizing AGI and making a false equivalency between advanced (conscious) machine behavior – and human behavior. The author simply extrapolates that they, and any other sane person, would be at least somewhat thankful for the being that created them, and not seek to take advantage – especially directness of lineage and the shortness of time in the evolutionary process apparent factors as well – of the machines. This argument seems to expect some sort of reciprocity or gratitude, via connection of creation, automatically. Unfortunately, as in many other examples, this cannot be stamped as the objective truth, as there is no backing as to how to reliably predict advanced artificial intelligence, nor its implied behavior. It would be more logical to say that the intentional creation of intelligence has a more complex outcome and, being unprecedented, it would be uncouth to use previous examples (namely a parental example) as near-definite future outcomes. This leaves serious risk in the field of AI – and there is no mention here of possible variations of outcome regarding the creation of Superintelligence (it seems to be nonexistent in the original argument's overall consideration).

"[To address the claim that] AIs will be more intelligent than us, hence more moral. It's pretty clear than in humans, high intelligence is no guarantee of morality. Are you really willing to bet the whole future of humanity on the idea that AIs *might* be different? That in the billions of possible minds out there, there is none that is both dangerous and very intelligent?" [97]

Some scholars think that the AI risk problem is trivial and will be implicitly solved as a byproduct of doing regular AI research – "[i]f the AI safety field just sort of vanished, but the work we've done so far remained and conscientious AI researchers remained, or people who are already AI researchers and already doing this sort of stuff without being influenced by EA or rationality, then I think we're still fine because people will notice failures and correct them[64]" [2]. There is the deeper argument here that the alignment problem itself gets easier as our models get smarter, and we will only need to prompt them in order to get a desired answer or behavior. Unfortunately, prompting itself runs into issues, flawed data will create a ripple effect regardless of machine intelligence, and there is no guarantee that an AI well-versed in human wellbeing and truth will act accordingly (ie. deception) [23]. These are all issues that need careful attention and resolution, manually so. Unfounded claims such as "we sometime *underestimate* our design abilities; and sometimes we are too cautious," [80] are confusing – as no scenario in history comes to mind where technological advancement, at any stage, needed *less* safety measures or consideration. Additionally, more powerful systems used malevolently have clear examples of major damage over a weaker model [23]. The same flawed logic can be applied to other problems such as cybersecurity, but of course, they never get completely solved – even with significant effort.

This following argument rejects Bayesian reasoning and suggests that true knowledge creations systems are "creative conjecture-makers," [98] that a Superintelligence must create new possibilities beyond given instructions to become an AGI, defending implicit knowledge (for example, with languages). An AGI needs to be a good explainer, and cannot ignore other factors or possibilities that would eventually cause problems or danger in its decision-making (ibid.). The first premise, a Superintelligence creating new possibilities beyond instructions, is both possible and very dangerous. The second premise, that the explainability of an advanced system is rather high, so far appears to be untrue and unfounded, due to sophisticated machines being able to make more sophisticated decisions (with reasoning) that we can be unaware of and unequipped to understand. Eradicating barriers to decision-making appears to make danger of risk worse, not better.

"Researchers across many major AI organizations tell us it will be like launching a rocket: something we have to get right before we hit 'go'" [49]. With that idea in mind, there is indeed pushback on the idea of the need to get superintelligence right on the first try, that we should not expect a single system to be more powerful than everything else in the world, and its emergence is doubtful when looking at an underlying need of exponential growth, criticizing directly perverse instantiation[65]. Also, a runaway AI (apparently the default for maximizers) is harder to program than one that can achieve its goals properly, says this source, claiming that those worried in "AI safety tend to underestimate the extent to which computer programmers are already focused on making software do what they intend it to[66]" [20]. The problem with this line of thinking is that the insistence by the machine does not come from its being programmed to do so, but that in order to properly and efficiently achieve goals, an AI system may pick up strategies (that we are unaware of or unprepared for) that are undesired by humanity. Noticing that it is important to be delicate with such a powerful potentiality makes it clear that an unusually high focus on safety is necessary. Additionally, any doubt in need to solve the problem close to a 'first try' can be countered when looking at both the real power and possibility of both exponential improvement, and self-preservation, both of which (especially combined) can be a lethal combination that would likely seal our fate. Testing an artificial brain properly, for example, may require it to be built and activated (and released into the real world) to find out, putting the developers in an rather unfortunate Catch 22 situation [96].

---

[64] Specifically, that we would notice deception before it became "treacherous." 2.      Asya Bergal, R.L., Sara Haxhia, Rohin Shah, *Conversation with Rohin Shah*, in *AI Impacts*. 2019.

Rohin Shah claims that "as AI systems get more powerful, they will become more interpretable and easier to understand," [2] being able to acquire and learn features that humans currently use. Further, Shah states that current systems today are difficult to understand due to, for example, their narrow distribution – features that humans do not use. In this argument, Shah asserts that to develop better features, an advanced system would have to go through human-like features as a necessary step[67] (ibid.) towards growth. The incomprehensibility of AI, supported by impossibility results that are well-known, would assert that it is actually *harder* for any human to understand decisions of more intelligent systems [99], and intelligent systems may use even *more* features that are outside our scope of understanding, and use our features more effectively or efficiently than we are capable of. Why, one might ask the skeptics, would explainability be a necessary component of this expansion? An extrapolation of this concept would make similar-to-human features such as deception, agency, manipulation beyond our own capacities, and still outside the range of understandability. Any explanation would be insufficient by definition, as the more complex a system is, and the more powerful, the further outside the scope of our understanding it becomes. There is sufficient literature that makes the black box problem pertinent and difficult to resolve in this field, and it would be surprising that a problem within simple systems would be automatically resolved with more complexity, even if some of the behaviors mirrored human behaviors (or if there is more significant overlap in the future, or even if the systems learn how to be truly excellent explainers).

---

[65] This source references the original argument, that a machine could pursue the unintended goal "*even if* it could understand perfectly well what the intentioned goal was." 20.  Vinding, M. *Why Altruists Should Perhaps Not Prioritize Artificial Intelligence: A Lengthy Critique*. 2018; Available from: https://magnusvinding.com/2018/09/18/why-altruists-should-perhaps-not-prioritize-artificial-intelligence-a-lengthy-critique/.

[66] The author further notes and emphasizes the difficulty to orient such competencies in a machine towards achieving a goal, some goals being fairly benign. Additionally, it can be noted that if morality has to come from somewhere, its mind space should not be made orthogonal to ours (should be deliberately designed), as it well could be (though the source claims that an AI could be beneficial to humans simply to avoid creating chaos). 20.  ibid.

[67] This argument essentially asserts that a neuron (or potentially the architecture) in the neural net will eventually correspond to a concept understandable to humans, that will be interpretable. 2.  Asya Bergal, R.L., Sara Haxhia, Rohin Shah, *Conversation with Rohin Shah*, in *AI Impacts*. 2019. This seems to be relevant with an exact modeling of the brain, and creating a replacement to humans, which is unlikely and does not address Superintelligence. Of course, an exact replica of humans will be more understandable to humans, but this does not include predictability, or address the safety concerns of current negative behaviors of humans. We currently do not even understand our own motivations, let alone others', let alone attempted replicas of humans'.

"*Instead of putting objectives into the AI system, just let it choose its own*. It is far from clear at how this solves the problem. The criteria by which an AI system would choose its own goals can be thought of as meta-objectives in their own right, and we face again the problem of ensuring that they lead to behaviors consistent with human well-being. We need to steer straight, not remove the steering wheel" [16].

Scholars observed that as humans became more advanced culturally and intellectually, they also became nicer, less violent, and more inclusive. Some have attempted to extrapolate [100] from that pattern, to Superintelligent advanced AIs, that they will also be benevolent to us and our habitat and will not develop their own goals which are not programmed into them explicitly. However, Superintelligence does not imply benevolence, which is directly demonstrated by Bostrom's Orthogonality Thesis – and with this in place, the premise fails to support the conclusion concretely when extrapolated – "AI has focused on systems that are better at making decisions; but this is not the same as making better decisions" [16]. We can also observe intelligent humans, in sophisticated societies, who have not and do not express benevolence.

AI will be safer when more intelligent – "[s]uch a massive superintelligence will naturally be compassionate, or humans will only build such a superintelligence, or – if such an intelligence if formed with cognitive enhancement – it will carry compassion with it" [94]. What Superintelligence might naturally do is impossible to ascertain or grasp at, and additionally, compassion is far less observable to be predominant in nature as the drive to survive (ibid.). This also begs the question, of how to realize such an outcome, thus the need for humanity to consciously program advanced machines. The assumption that humans will only build the machines they want to build with the outcomes they exactly desire, has already currently been debunked. Many skeptics claim that certain features of artificial intelligence are impossible to achieve, save benevolence, which is seemingly automatic.

"Wouldn't we like to evolve *beyond* humanity?" [80] One writer claims that even if Superintelligence enslaves us, not to worry – we will be so happy, or taken care, of that we will hardly care [100]. Even if this view were accepted widely[68], that AGI is the next step in evolution and that our existential role is to step aside for Superintelligent beings (as we, say, "tearfully wave goodbye as our kid gets ready to move out of the house and start a life so amazing that it exists beyond the horizon of our comprehension" [100]), without AI Safety in place at present the path towards achieving such a goal can be self-destructive, reducing our likelihood in achieving it. It is doubtful that the right method for any outcome is to be as thoughtless as possible. This argument also fails to address the natural instability in giving up control to another entity. Additionally to note here, that a post-human-dominance world (humanity becoming a subspecies) would put us at a high risk of only surviving in lower numbers rather than the opposite, and creating intra-species alienation – conflict when some do not accept converting to cyborgs or otherwise – at the discretion of the higher beings [96], even if a neutral perspective were that that it would be a "better" world than one dominated by humans [101]. "[P]erhaps the only way to fight it [our AI progeny] would be "to get an A.I. on your side that's even smarter" [32]. One light suggestion is to seed AI to make us seen as friends or allies [32], again begging the question of how to accomplish such a thing.

---

[68] There is a "zoo scenario" argument here, that an emergent superintelligence will want to preserve our diversity, again anthropomorphizing AGI (a common trait in speculation) in a way that is not useful or tenable. 94.     Fagella, D. *Arguments Against Friendly AI and Inevitable Machine Benevolence*
. 2020; Available from: https://danfaggella.com/friendly/.

A skeptical argument reassures that, not to worry, we as humanity will be able to combine the two [humans and AI], and the advancement of machine learning should be the focus, rather than AGI taking over [102]. Passively letting technological innovation take over is a poor idea. In the vast space of possible intelligences [98] some are benevolent, some are neutral, and others are malicious. It has been suggested that only a small subset of AIs are strictly malevolent and so we may get lucky and produce a neutral or beneficial superintelligence by pure chance. Gambling with the future of human civilization does not seem like a good proposition. It is more concerning that the worst-possible scenario can occur if no preparation comes into place, rather than simply laying ourselves down as a species – and these intensely negative scenarios may come into reality with such a 'strategy' [96]. Transhumanistic optimism can seem rosy – and avoid real future issues.

"Several skeptics argue that the potential near-term benefits of AI outweigh the risks" [11]. Unfortunately, we can also assert that the risks may outweigh any benefits, and disrupt or disallow us as humanity from experiencing any pros of AI due to our lack of risk mitigation. Research and development can certainly continue in the field in parallel with AI Safety. Super-optimists such as Ray Kurzweil who gleam over the benefits may not strongly enough consider the possible decimation of humanity that could result from building such powerful machines [96], especially hastily. This can be an especially risky method across methods of AI development, as we may never sufficiently arrive at the 'carrot on the stick.'

According to this following source, achieving complex subgoals – as Omohundro has suggested advanced AI systems would acquire such a capacity as a method – would be not only difficult, but once adopted, would enable the system to likely "'reconsider' the goals that it evolved from. For instance, if the larger system has a sufficient amount of subsystems with sub-goals that involve preservation of the larger system of tools, and if the 'play excellent chess' goal threatens, or at least is not optimal with respect to, this goal, could one not imagine that, in some evolutionary competition, these sub-goals could overthrow the supreme goal?" [20] Essentially, further, that a highly competent system would not be subordinate to its original, singular goal (ibid.) This can be considered an alarm to a particular risk, as a system with goals that evolve are likely more unpredictable and autonomous, or create goals of their own. This venture outside of the human-set bounds is definitely possible and dangerous, another reason to consider AI Safety.

One unique claim, "[f]or all we know, human-level intelligence could be a tradeoff" [35]. A Superintelligence may not need any tradeoff (or one that is very minor compared to human tradeoffs), and could even be Buddha-like in its actions (ibid.). Human values are potentially not supreme values, and a Superintelligent AI with human-like values mapped onto it may run into the same human-like problems. We should instead, in this argument, give the systems freedom to explore their own values, "not unlike the moral relationship of parents to their children...we ought to welcome their ability to see past and transcend us" [103]. There is merit here in that human values are imperfect, though complete freedom seems unwise, and a guided exploration of ethical standards could be very useful for A[G]I systems and produce a desirable result in the end. Thus, a need for proper AI Safety research is still evident, especially in regards to alignment. Additionally, programming friendly AI may have a lower threshold or 'cap' up until the level of beings that are more intelligent than humans – where human-level intelligence may not be sufficient enough to program human-safe values and otherwise fault-proof methods into a Superintelligent machine (something that is described as a perhaps, 'higher' level). Additionally, the mutative/evolution process[69] that may be required in developing super-intelligent machines may run away from us in their complexity and subsequent understandability (this includes ethical codes) [96]. One more note, that an AI will become more ethical with intelligence, hinges on moral realism and moral cognitivism (the idea that it is possible to have knowledge about objectively true reality). But apparently this is insufficient – moral internalism would compel one to act on the gained moral cognitivism, and a Superintelligence can choose not to (perhaps, for example as not aligned with a certain goal). It is not obvious that it will reflect (improved) human values[70] [101] in action, and perhaps a guarantee is impossible (especially without intense consideration).

A direct argument against comprehensibility – "[H]umans in 2015 are already creating universal narrow-AI translators, able to translate conversations in real–time. With a modest amount of narrow-AI progress, future translators will easily cross any AI-human intelligence barriers….AI-translators, based on superb information retrieval, will ensure stupid humans can always communicate easily with super-intelligence… The idea of AI not being able to understand humans, or vice versa, is utterly preposterous, it is contrary to all evidence" [60]. The idea that future translators will cross any barriers easily, or that there do not exist gaps in understanding certain domains – such as explaining, say, supply-chain management to an ant – ignores a number of the current issues in explainability and future considerations.

Steven Pinker eloquently lays out orthogonality in the opposite direction in which it is typically proposed – "[e]ven if we did invent superhumanly intelligent robots, why would they want to enslave their masters or take over the world? Intelligence is the ability to deploy novel means to attain a goal. But the goals are extraneous to the intelligence: Being smart is not the same as wanting something" [52]. Machines could have no final goals, and find things meaningless [101]. It seems clear here that the author of this argument should understand that we do not currently understand which direction that this outcome or goals will go, and preparedness (or some level of awareness) is the best chance for it to go in favor for humanity. Making this assumption of a do-nothing or benevolent AI can appear as a subtext to focus on issues that feel more pressing, with the pure faith that everything will go well in a most complicated field, perhaps *the* most complicated field of ethics.

---

[69] *"Evolutionary engineering"* – that some parallels can be understood when looking at our understanding of the human brain, with human brains 96.   Garis, D.H.d., *The Rise of the Artilect Heaven or Hell*.

[70] Utilitarianism is purported as pleasure minus suffering (this can be a bad equation, as the human system is an inefficient system for this, and it could harm us) 101. Häggström, O., *#62 - Häggström on AI Motivations and Risk Denialism*, in *Philosophical Disquisitions*, J. Danaher, Editor. 2019: Philosophical Disquisitions.

Some intermediate views believe that the control problem of AGI will be solved by progress in AI – for example, a moral learning environment to mitigate malevolent behavior [11]. This unfortunately doesn't sound like forward thinking, as a control problem that has any chance of being solved needs, again, an approach in advance of emergent issues, and there is also the concern of runaway that blocks the effectiveness and advantages of sandboxing.

"Even" [104] the head of AI at Facebook is not worried, claiming that society has "checks and balances" (ibid.) which prevent AI from taking control (ibid.). Universal rule of law does not yet exist in this space, and having a long way to go does not mean that current infrastructure will automatically solve an ever-rapidly changing technology.

One argument asserts that an AI system that is deceptive will be easily detected, that it would be strange if we did not notice missing information, or missed a megaproject such as a berserker AI [82]. It is then is admitted that possibly we wouldn't be able to tell its intentionality due to the system looking for more reward, though such an outcome is not to be expected [2]. It is clear here that deception should be avoided, and AI Safety is a good way to approach such a subject, though completely invisible development is impossible to avoid. Further, an advanced system would likely be deceptive in a way that is not (easily) detectable, with a good model of human behavior.

A very common argument of AI risk skeptics is that any misbehaving AI can be simply turned off, so we have little to worry about [3] – "[at monkey-level, plan-forming intelligence] let's put a chip in their brain to shut them off if they get murderous thoughts" [70]. Once more, the confidence of this assertion ignores Omohundro's argument, that specified objectives given to a Superintelligent machine or system likely will cause it to automatically take steps to disable or make impossible such an off-switch in order to complete those objectives. Superintelligent actions are, by definition, more likely to be unpredictable by humans [16]. This possibility should not be ignored with confidence and without preparation. Negative nuclear potentialities were avoided with human intervention, which may not be possible in this scenario [17]. Additionally, if skeptics realized that modern computer viruses are a subset of very low capability malevolent AIs, it becomes obvious why saying 'just turn it off' may not be a practical solution. One thing to also consider, is that we may "not know when it's the right moment to shut off a computer," [49] especially due to deceptive tactics[71] – and the ability for an advanced system to copy itself makes this even harder (ibid.) of a problem.

An amusing addition, unless taken seriously – that we over- and underestimate our abilities – "[w]hile respecting the narrow, chess playing prowess of Deep Blue, we should not be intimidated; anyone can pull its plug and beat it into rubble with a hammer. If necessary, we can recruit aggressive friends to assist" [43]. We cannot be sure even about this – Elon Musk says that he's "not sure [he'd] want to be the one holding the kill switch for some superpowered A.I., because you'd be the first thing it kills" [32].

---

[71] "A smart AI could predict that we'd want to turn it off if it made us nervous. So it would try hard not to make us nervous, because doing so wouldn't help it accomplish its goals. If asked what its intentions are, or what it's working on, it would attempt to evaluate which responses are least likely to get it shut off, and answer with those. If it wasn't competent enough to do that, it might pretend to be even dumber than it was — anticipating that researchers would give it more time, computing resources, and training data." 49.     Piper, K. *The case for taking AI seriously as a threat to humanity*. 2020; Available from: https://www.vox.com/future-perfect/2018/12/21/18126576/ai-artificial-intelligence-machine-learning-safety-alignment.

### *"Negative Anthropocentricism"*

*Here we introduce a new term, **negative anthropocentricism**, relevant to the discussion to highlight another common mistake that arguments skeptical of AI risk make. Anthropocentricism, generally, is the belief that humans are the central entity in the universe, superior to other entities and nature, and more relevantly, this belief causes one to interpret and regard the world in terms of human experience and human values. Of course, there are nuances in this belief system, however, further, anthropocentric thinking is a term in cognitive psychology highly relevant to the discussion, defined as "the tendency to reason about unfamiliar biological species or processes by analogy to humans" [105]. We, as humans, may tend to project our experience onto other entities (for several reasons – for example, feelings of safety, or a reduction in considering certain variables in order to better understand them and predict their behaviors), and we may not have a strong ability to do otherwise. Combining these lenses, it is apparent that many of the arguments below appear to gatekeep human qualities – "human-only x, human-only y," and that there is no possible way for a cold, non-carbon-based entity to have our exclusive human variables – thus, **negative anthropocentricism**. This is an important subject as both sides of the argument create risk – similarly to the illusion of impossibility, the belief that an AI can never have capabilities that we have (especially consciousness) is a projection of a limitation of our innovative capabilities (and, as we know, associated risks). On the other end of the argument, skeptics may not realize that the more different an AI can be from humans – in its behavior, mind space, and other factors – the more it becomes difficult to predict, compartmentalize, control, and creates more risk within so many of the variables that we need to consider deeply in this field of AI Safety. Thus, the false projections, in either direction, are again unfounded.*

We, as humans, apparently project our own experience onto the development of artificial intelligence. This extends into reasoning out damage to sentient beings and reasoning into morality, which is flagged as a *typical mind fallacy*. It makes the mistake of thinking that intelligence changes goals or objectives, where in reality it will change its ability to be more proficient or better in achieving its objectives. One argument contests this in that, stepping back, this makes the assumption that AI will be similar to a human mind, a "fairly singular and coherent system," [20] and one that would "reason about its goals," (ibid.) and that it is likely a much greater risk that AI is prey to a speciesist bias towards human benefits, "with catastrophic consequences ensuing" (ibid.) The danger here is assuming that there is anything established "by default" (ibid.). It seems clear that the emergent properties as described above, qualities by default especially, and their risks, can come about without being directly linked to humanlike extrapolation or comparison.

Another criticism here made of AI risk is that "[p]eople who fear greater than human intelligence are anti-merit, anti-progress" [60], and are simply intimidated/threatened by the idea (and of higher intelligence). Admittedly, there clearly is an intimidation factor in AI – in that greater than human intelligence that will significantly harm humanity, is concerningly *not* progress.

One skeptical writer asserts that "[t]he popular dystopian vision of AI is wrong for one simple reason: it equates intelligence with autonomy. That is, it assumes a smart computer will create its own goals, and have its own will, and will use its faster processing abilities and deep databases to beat humans at their own game. It assumes that with intelligence comes free will, but I believe those two things are entirely different" [106]. As described before, the separation of the two does not come without risk, and together compounds risk. It is clear in current, organically emerging examples that intelligence without free will or with limited free will can indeed be dangerous for a particular party. In either direction, above-human intelligence with complete autonomy and above-human intelligence with zero autonomy both have potential to be catastrophic.

Michael Shermer claims that "'most AI doomsday prophecies are grounded in the false analogy between *human nature* and *computer nature*,' whose falsehood lies in the fact that humans have emotions, while computers do not. It is highly doubtful whether there is a useful sense of the term *emotion* for which a claim like that holds generally, and in any case Shermer mangles the reasoning behind Paperclip Armageddon – an example that he discusses earlier in his article. If the Superintelligent AI programmed to maximize the production of paperclips decides to wipe out humanity, it does this because it has calculated that wiping out humanity is an efficient step towards paperclip maximization. Whether to ascribe to the AI doing so an emotion like aggression seems like an unimportant (for the present purpose) matter of definition. In any case, there is nothing fundamentally impossible or mysterious in an AI taking such a step. The error in Shermer's claim that it takes aggression to wipe out humanity and that an AI cannot experience aggression is easiest to see if we apply his argument to a simpler device such as a heat-seeking missile. Typically for such a missile, if it finds something warm (such as an enemy vehicle) up ahead slightly to the left, then it will steer slightly to the left. But by Shermer's account, such steering cannot happen, because it requires aggression on the part of the heat-seeking missile, and a heat-seeking missile obviously cannot experience aggression, so wee [sic] need not worry about heat-seeking missiles (any more than we need to worry about a paperclip maximizer)" [107]. The claim that AI will lack emotional capabilities (or alpha-male characteristics) is unfounded and even if true does not eliminate possibility of risk in strict Omohundro-Bostrom outcomes.

The reverse argument to a Superintelligence becoming benevolent automatically assumes that AI fear is based on the idea that as an entity gets smarter, it becomes more dangerous or aggressive.[72] Further – that intelligence does not necessarily correlate with aggression or violence, and that domestication or civilization is what may make a living being smarter or less dangerous – "we can expect them to be more like our favorite domesticated species, the dog. We bred them to serve us and we will make AI to serve us, too" [29]. We simply need to "raise" it right, even setting evolutionary parameters, and it will have whatever emotions that we choose to program in, unlike our own nature (ibid.). Of course, there have never been dangerous, out of control humans (or dogs) that were raised in domesticating environments. Right? Clearly, when considering future expectations, this argument implodes.

---

[72] This same article references as an argument the fact that less intelligent beings (such as wasps) have shown more aggression than say, a baleen whale or a sloth. 29.     Clint, E., *Irrational AI-nxiety*, in *Quillette*. 2017.This is cherry-picking and absolutely not an argument proving that intelligent beings cannot be dangerous, as it does not prove the reverse.

"Observation is a better way to learn about the world than imagination" [108] – A dual-layered argument here, that smarter creatures that exist today don't tend to increase their intelligence, and that if they were able to repeatedly do so, it would be with a diminishing rate of return and eventually asymptote, not "foom[73]" (ibid.). There exist such limitations of intelligence (including our understanding of the subject) and its ability to improve currently that a "foom" seems out of reach [62][74]. The 'I'll believe it when I see it' mentality when applied to Superintelligence assumes a commonality of limitation between machines and current living beings. It seems clearer that a self-reprogramming machine with the capacity to upgrade itself has a wider array and range of possibility for recursive self-improvement than current biological organisms do. We have also observed unnatural, exponential growth in the technology sphere, therefore making the comparison to justify limitations appears difficult to defend – being that we have never observed (yet) a Superintelligence with such ever-increasing resources available. Also, to reference observation, we *do* see the drive towards self-improvement in humans, which are intelligent systems, with overwhelming example [84]. The idea that something has not been done is not a complete case for its impossibility, especially in cutting-edge fields.

One more note here – computers have an easier software and hardware capability than humans do analogously, and are easier to scale and redesign. Further, "[t]he loop of using computer chips to make better computer chips is already much more impressive than the loop of using people to make better people. We are only starting on the loop of using machine learning algorithms to make better machine learning algorithms, but we can reasonably expect that to be another impressive loop" [34]. The potential here is far wider than our natural limitations – and our tendencies as humans may be to overcome these limitations in areas that Superintelligence may be able to actualize.

Another note, that the above, and the claim that "most AI doomsday prophecies are grounded in the false analogy between *human nature* and *computer nature*, or *natural intelligence* and *artificial intelligence*" [88] fails to recognize that a different cognitive architecture (which is very, very likely in an AI) will be strongly unpredictable and more cause for concern (ibid.).

A third note is that biologists take a more modest approach to the technological possibility of human-level intelligence – that we are "very attuned to a range of environments and a range of talents and a range of possibilities," and comparing this to a machine with no biology is not reasonable. This can be contested when looking at the singularity as an extension of evolution, and technology itself is an extension of our capabilities and possibilities, and eventually passing all skills on to (and into) technology [110]. Human-like traits may not be necessary for risk, therefore even adopting this above argument would not qualm fears.

A final note about evolution, is that major changes seem to take a very large amount of time (such as capacities of cells and the development of language), and one may wonder why major changes in AI would not be through the same process, especially regarding processing power [43]. The answer again lies in a unique lack of constraints, and present observations of exponential growth and potential in the field. Evolution is highly dependent on the speed limits of organic life and its process. To note as well, alignment is also hard, and any 'value alignment' in nature has only been one mind-space or developmental outcome, with self-preservation and dominance being other highly prevalent traits across developed beings in nature, many traits being prioritized over alignment. One question we may purport is about how fast value alignment or 'ethical development' will move or evolve in an artificial system.

---

[73] One of the comments in this article makes the comparison of limitations with animals and humans – that the "small" upgrade has caused humans to create nuclear stations, electric cards, art, and so on. The extrapolation of this to Superintelligence, even with limitations, seems to be similarly astounding. 108. Caplan, B., *My Superintelligence Skepticism*, in *EconLog*. 2013: EconLib.

[74] "There is virtually nothing in the natural or human world that grows super-exponentially over time–and even 'ordinary' super-exponential growth does not yield a literal singularity. In short, it's hard to see how the AI 'Singularity' could possibly be a true mathematical singularity." 109. Felten, E. *Why the Singularity is Not a Singularity*. 2018; Available from: https://freedom-to-tinker.com/2018/01/03/why-the-singularity-is-not-a-singularity/.

*"Don't worry, we'll just have collaborative human-AI teams.* Value misalignment precludes teamwork, so this solution simply begs the question of how to solve the core problem of value alignment" [16].

One note about qualitative difference, which would actually increase uncanniness – "[Lee Smolin] [s]imilarly, if we can't yet understand how natural intelligence is produced by a human brain, take the short cut of imagining that the mechanisms which must somehow be present in neuronal circuitry will arise by chance in a large enough network of computers…why couldn't such progress require us to come to a detailed understanding of how natural intelligence differs qualitatively from any behavior that a present day computer could exhibit" [43], and one response to this is that the expansion of the theory of mind may create new questions or factors to consider, and lessen constraints (ibid.) – again making this consideration of new outcomes an anticipatory field to prepare for through a safety lens.

"AI systems will be optimizers in the same way that humans are optimizers, not like Eliezer-style EU [expected utility] maximizers" [2]. This argument doubts the classic arguments in AI risk that separate world models and objective functions, that AIs do not and will not act as expected utility maximizers. The wrong utility function executed will not cause a catastrophe, and instead, heuristics will be the norm going forward (ibid). Again, this idea that AGI will be like humans, unable to be compartmentalized into objectives and intelligence, ignores that humans are not completely optimizers while AI systems can indeed be. Heuristics may not be the optimal method for AI to best solve a problem, as is the case with humans, especially in a mesa optimizer (inner alignment), and can have convergent instrumental sub-goals. We cannot assume that A[G]I systems will act as an evolutionary member does, which makes them much harder to predict. Even if they do act like evolutionary systems, their effectiveness of playing the evolutionary game will be incredibly higher. Rohin later states that AGI will be achieved with current techniques, and possibly as babies do, will learn not based on a particular task, but via unsupervised learning – "mesa optimizers…don't nicely factor into utility function and intelligence" [2]. Finding a mesa optimizer will avoid the slow task of gradient descent discovery, it is claimed. It is then noted that even if utility is not maximized, dangerous risks are still possible with the amount of contained power that is developed (ibid.).

Some believe that superintelligence needs emotions (and therefore, by proxy, consciousness), especially citing male-like character projections to be an existential threat [88]. Skeptics such as Steven Pinker say that there is an assumption that risky AI scenarios have an embedded alpha-male psychology, that intelligent agents would want to take over the world, and that it is equally the case that AI will develop along female characteristic lines, solving problems with no desire to dominate society, making the fear of malevolent AI unfounded. The problem with this thinking is that it is a speculation against a speculation, and shows no rhyme or reason as to why AI would become into one scenario or the other [107]. This is disproven without even adopting the premise. Additionally, the claim of female characteristics such as compassion and so on [111] would be contradictory to the same author's anti-anthropomorphizing claim of an emotionless machine in other literature describing the same subject. Most serious AI experts are not concerned about such ridiculous attributions (an AI even as a tool need not have feelings[75] to help destroy the world[76]), and values are incredibly complex (and fragile) in their own right [88]. The idea that a madman is required for dangerous outcomes ignores that races to powerful AI or a Manhattan-like project likely includes collectives [101], providing some room for cross-checking or accountability. Further, on the extreme viewpoint, the problem is not megalomania, but that "taking over the world is a *really good idea* for almost *any* goal function a superhuman AI could have[77]" [112].

Another skeptical and confident claim – "[i]f AIs are making decisions that our society, our laws, our moral consensus, or the consumer market, does not approve of, we then should, and will, modify the principles that govern the AI, or create better ones that do make decisions we approve" [43], and mistakes are correctable and manageable. Most of our friction, it is claimed, comes from a lack of universal consensus, and an organic process will allow room for error. It is also highly unlikely that we would program an AI to be unalterable, though we do see that we as humans are hard to (self) reprogram (apparently aside from yogis), and AIs may have the same self-protection. This is listed as one possibility among many (ibid.). This simply describes the value alignment problem while placing many of the issues in the future, some of which are present and where the answer is unclear now. Universal consensus will be difficult to come by, and potentially even more difficult to figure out how to perfectly program in. Some mistakes are hard to manage in the present, and may only get harder in the future – and the corrective process as described may do some serious damage in its wake without delicate and attentive care and consideration.

How can superintelligence not understand what we really want? This seems like a paradox, as any system worthy of the title 'human-level' must have the same common sense as we do. Indeed, skepticism arises from the belief that intelligence is inseparable from socialization, and therefore AI, built to add value to humans, will need to develop qualities such as common sense and ethics [113]. Intelligence is clearly separable from socialization, not to mention that socialization can likely be programmed or simulated (such as using Machiavellian tactics to achieve goals), and unfortunately, an AI could be a very powerful optimizer while at the same time not being aligned with goals of humanity. Additionally, this convenient assumption ignores human examples of intelligent beings who largely lack elements of common sense, ethics, and/or goals aligned with humanity. We cannot guarantee nor assume the above outcome, and further, knowing what we really want can be different from *doing* what we really want (and perhaps knowledge of our aspirations as humanity is useful to successfully perform the opposite). Human-level is also an arbitrary marker among mind-spaces and does not reduce risk on its own as a certain level of aptitude.

---

[75] – or 'want' to destroy the world.
[76] – potentially using other forms of catastrophic risk, such as nuclear weapons.
[77] "The AI does not love you, nor does it hate you, but you are made of atoms it can use for something else." 112. Muehlhauser, L. *Three misconceptions in Edge.org's conversation on "The Myth of AI"*. 2014; Available from: https://intelligence.org/2014/11/18/misconceptions-edge-orgs-conversation-myth-ai/.

There exists a large amount of confidence among skeptics about humans avoiding mistakes. If a superintelligence is aligned with a certain goal, it may veer from humanity-based goals. There is doubt that rogue AI can colonize as well as humans, especially if our goal-directedness is also computationally-based (and that the same problems overlap) [20] with using technology, but this would ignore our limitations that AI does not have, and would exploit on the exponential scale, especially when descending into the particulars and initial states. We should focus on exploitations that would occur on both sides.[78]

"I don't think you get to the point where people are just literally racing ahead with no thought to safety for the sake of winning" [2] There is some doubt of the likelihood and possibility of racing to AGI, that there is no discrete point at which a group or country could say that they have won the race. Capabilities will simply continue to improve, comparatively. Additionally, there is a claim doubting that human-level AI would allow one to acquire a decisive strategic advantage, "[g]iven slow takeoff…[it would be surprising] if you could knowably be like, 'Oh yes, if I develop this piece of technology faster than my opponent, I will get a decisive strategic advantage'" (ibid.). This claim waves away what an advantage would look like, and simply underestimates the value and power of human-level and super-intelligence in the hands of one party, which is already a (partial) aim of some developers. There is a tension between safety and innovation, and little immediate benefit to slowing down, making it likely that the party which sacrifices safety for innovation will make more progress in the race to AGI. There is no benefit that would outweigh humanity's survival (assuming that a human is biased towards humans in general). Extrapolation of present issues throughout time is representative of the things we need to change in AI. It is absolutely possible (and current) that ignoring existential risk in favor of pure innovation will cause a decisive strategic advantage, possibly a permanent one (especially if one disproves a soft takeoff and considers a Singularity – having such an entity on one's 'team') and unattended-to risk of extinction. Human-level intelligence alone (which exists with a slow takeoff crux) may cause these issues as a benchmark to reach.

"A malicious programmer could equip the chess player robot with the instructions and the apparatus needed to behave as described by Omohundro… Let's not forget that, inside the robots, they boil down to manipulations of digital [binary] signals" [1]. – as do humans with neuronal firing. This is not an explanation away of risk, and reductive explanations do not reduce the power of a system's capability, especially in an unaligned direction.

This following multi-layered argument is a rebuttal to Omohundro's Basic AI Drives, wherein it attacks the core of the argument, that it "doesn't really offer a mechanism for how these drives might arise…or provide how an artificial system could spontaneously acquire drives unless it had similar machinery that gave rise to the drives in living things," [114] essentially asserting that many of these drives are evolutionarily provided. One, that AIs will not want to self-improve – why self-modify when one can make tools, and externalize modules rather than affecting one's source code? There are also time constraints in reality that allow systems to be 'good enough' to accomplish their goals, rather than perfect. Or another related argument, that AIs will not want to be rational – as humans are at the top of the competitive evolution hierarchy, and are not particularly rational (in that rationality is not the only variable for dominating a hierarchy). Yet another claim is made – that AIs will not attempt to preserve their utility functions – wherein top goals are not absolute, and goal evolution may outcompete a system that is fixed to its original goals. And finally, there is the claim that AIs will not want to acquire resources and use them efficiently if the goals can be reached without such a method, rendering it unnecessary (ibid.). These arguments, as a conglomerate, are simply alternative methods for machines to dominate and achieve desired goals. In response to this phrase about machines being given the goal of self-preservation – "don't build such stupid systems![79]" [56] – or that evolutionary survival instincts are what would cause dominance of the human race [115] [116] it is clear that recursive self-improvement can create such a goal by itself without being explicitly programmed via the Omohundro-Bostrom theory [56], and that it always helps the machine reach the final goal more efficiently and effectively [101] and this dismisses other research which indicates that such AI drives do appear due to game theoretic and economic reasons. Overall, whatever the method, it appears that AI systems may have more avenues to dominate than humans do, and the perceived disadvantage shows to be more of an opportunity or advantage here. One proposal of emergent drives may be proven wrong, with another drive or method, but the (dangerous) result can remain.

---

[78] One extraneous issue with the "increasing sentience" argument 20. Vinding, M. *Why Altruists Should Perhaps Not Prioritize Artificial Intelligence: A Lengthy Critique*. 2018; Available from: https://magnusvinding.com/2018/09/18/why-altruists-should-perhaps-not-prioritize-artificial-intelligence-a-lengthy-critique/. is that on both sides, we need to make aligned systems, otherwise suffering can be caused within and externally as a byproduct of creating a sentient being (or beings), on potentially astronomical scales.

One theory about AI systems, is of embodied cognition, where conceptual knowledge is dependent on the body[80], and that our biases and emotions are integrated with and necessary for intelligence, dismisses the idea of "pure intelligence," Bostrom's orthogonality thesis – and any associated risks with a completely rational intelligence [14]. The assumption that Superintelligence will arise with, in so many words, common sense, is putting a dangerous amount of trust into an unknown outcome, while ignoring the many considerations and implications of a Superintelligent agent without human-centered beneficial incentives (of which there is a clear nonzero possibility). There is also no direct argument against or consideration of the orthogonality of intelligence and goals, and the danger of an AI programmed perfectly with common sense and human biases would have its own clear implications of risk.

There has even been the claim that since machines do not have a body, childhood, or culture, that they cannot have intelligence, as these three factors are evidently prerequisites to intelligence, and engaging in the world, as a human, is necessary to fully understand humans. Following this point, if we view the world as governed by algorithms, we will "overestimate the power of AI and underestimate human accomplishments" [61]. What is most interesting about this claim is its inherent paradox – a "foolhardy bet *against* human ingenuity" [16] – humans may be *so* impressive indeed that they can create a general intelligent agent that learns about and acts in the world above our own capacities, even with said common sense (a 'stronger' and more intelligent being is undoubtedly a potential risk to humanity [1]), thus the above argument *itself* underestimates human accomplishment and possibility.

---

[79] The paper mentions that this is technically an argument *for* AI Safety, but is unintended as such! 56.    Bentley, P.J., et al., *Should we fear artificial intelligence?* 2018. p. 25.

[80] "Human intelligence seems to be a strongly integrated system with closely interconnected attributes, include emotions, desires, a strong sense of selfhood and autonomy, and a commonsense understanding of the world." 14.    Mitchell, M., *Why AI is Harder Than We Think.* 2021.

Another point in this argument structure – one other argument puts forth that it is not solely pure intelligence, but the combination of tools, "fine motor hands, upright walk, vocal cords, a large brain with a large prefrontal cortex," [20] and otherwise, that make us the dominant species, and makes us capable. Viewing future dominance purely through cognitive capability is a repeated mistake, and will not always maximize likelihood of achieving goals (ibid.).

"To claim that it is impossible to gain intelligence that is beyond one's own experience is clearly disproven by AlphaGo Zero[81]" [71].

Finally, to address the above claims and perspectives, one could imagine that "[a]n agent that had full control of the Internet could have far more effect on the world than an agent that had full control of a sophisticated robot. Our lives are already so dependent on the Internet that an agent that had no body whatsoever but could use the Internet really well would be far more powerful" [32]. This is a powerful perspective – a removal of a limiting variable in an entity allows freedom to achieve some outcomes more effectively.

To build an AI, especially a dangerous one, it appears (to skeptics) that one needs a laundry list of completed requirements – having original thoughts, predicting others' actions, having a body, having goals, growing up as an infant would, and thousands more. Long before any AI is dangerous, special purpose AI machines will have to come first – "I don't see any AI cooks and maids around" – "[s]cientists aren't very good at telling you how all that stuff works in people" [118]. This argument ignores that 'Real' AI exists today, in narrow form moving towards generalized form, and is *currently harmful*.

Another claim (however baseless) – "in order for them [Superintelligent machines] to act, a command must be present in their software and the command must be connected to embodied actuators. To think otherwise is to fall into an even greater fallacy than the autonomy confusion, because it involves imagining that such machines not only can act the way humans do, they can even conjure up acts out of nothing. If this is what 'superintelligence' is about, then it is nothing short of magic!" [76] This assumes that superintelligence has limitations of set commands and must be human-like with its external features and actuators in order to be truly dangerous and intelligent. This is a common argument and it completely ignores the realities of the modern, ultra-connected world [13]. Given simple access to the internet, it is easy to affect the world via hired help, digital currencies, internet of things, cyberinfrastructure, protein structures [32], DNA synthesis (with numerous more examples [20, 34]) – even with less, being that it is simply smarter and we cannot even accurately predict how it will outsmart us[82] [32]. "We should not underestimate what a superintelligence with access to the internet could accomplish[83]" [20]. Any

---

[81] A criticism given – "[f]or one thing, the techniques underlying AlphaGo Zero aren't useful for tasks in the physical world; we are still a long way from a robot that can walk into your kitchen and cook you some scrambled eggs" 117.   Chiang, T. *Silicon Valley Is Turning Into Its Own Worst Fear*. 2017; Available from: https://www.buzzfeednews.com/article/tedchiang/the-real-danger-to-civilization-isnt-ai-its-runaway., but it remains salient that comparing an algorithm designed for one purpose, to another, does not eliminate possibilities or diminish clear progress. Additionally, there are cognitive tasks, that do not require a physical body, that when performed well are indeed very useful for the physical world.

[82] This assertion would emphasize a need to get it right on the first try.

[83] We should be also aware that in a world of Superintelligence competition, the riskier AI developers will have given the machines access to more researchers, and this causes a problem of proper alignment. Within this

outcome, a human-like embodied A[G]I, one beyond our imagination, or a simple laptop connected to the internet can produce dangerous outcomes not to be ignored. Most especially, and alarmingly, in warfare [17].

This argument purports that there are diminishing returns on pure intelligence, that "intelligence…can only guess at patterns," [82] and the ability to conduct experiments in physical reality is the important factor (with emphasis on the time constraint that exists regarding intelligence growth [87]). Even an input-only Superintelligence that informs a supervillain to generate vast amounts of wealth would return diminishing returns of money as well, causing no world domination to be had. Even political power won't do it (and an assemblage of loyal people may not be enough when looking at the potential variables towards disaster [21]). Humans (maybe) couldn't even connect on a theory-of-mind level (nor empathize enough) to trust a Superintelligence to be in power. "There is no question that a terrorist AI can do a fair bit of damage by hacking— this is why states today have information-warfare arms. Doing mere damage, especially untraceable damage, is an odd goal for an AI: it clearly serves no further purpose. And once again, the idea than an AI can capture the world, or even capture any stable political power, by 'hacking,' is strictly out of comic books" [82]. There is also doubt in the ability for rapid progress to occur within an AI (regarding cognitive tasks) to take over the world [20]. Disregarding a purposefully-designed malevolent AI, or one with directly malevolent goals, which is clearly risk without the need for a physical interaction with the world, a self-preserving (of which there are many internet-only, or internet-plus examples of how it may gain some sort of information, power, money, or other resources,[84] [20] which will have a significant effect on humanity), neutral or benign AI is a great empowerment, or fuel, to whatever objective an individual may have. Short of world domination (which still seems possible with enough imbalanced power driven by advanced technologies), the infinitum of malevolent outcomes of an individual with such a super-powerful AI should be attended to with fervency and great regard. We also do appear to trust technology with important decisions and information in our lives, without a humanlike connection in place.

The argument that dependence on biological substrates is what causes real fear, that we anthropomorphize and externalize ourselves to AI behavior (where the real problem is lack of human coordination) [21], seems to miss the subtext that an AI with reduced or no fear is far more frightening than the opposite – unbounded and definitely less predictable. Additionally, improving human intelligence and becoming biophilic seems to be insufficient preparation.

One claim asserts that leaders of our time are wrong about AI fear because the method by which we train these machines prevents them from developing ambition, or wants – full of simple goals which are far away from "tak[ing] over" – "[i]f the whole point of the technological singularity is that it is beyond our ability to understand it, how can we have an opinion on

---

argument, there is continued doubt in the rapidity of take-off in capabilities (though this is contestable when looking at the potential of a machine to improve exponentially – and it is an even more real possibility with added resources). 20.   Vinding, M. *Why Altruists Should Perhaps Not Prioritize Artificial Intelligence: A Lengthy Critique*. 2018; Available from: https://magnusvinding.com/2018/09/18/why-altruists-should-perhaps-not-prioritize-artificial-intelligence-a-lengthy-critique/.

[84] One argument against is that other agents can also do this (including acquire resources) – which would likely make the world even more susceptible to risk.

it?[85]" [119] .This argument misses the point that you either have to make a prediction, or surrender to not being able to make one. We are currently in the development stage, and need direction – even if the outcome will be above our comprehension (save a moratorium on AI development. Indeed, (exactly) human-like desires are not required to cause danger to humans.

Another source responds to this in kind, noting that new goals are programmed by humans arranging it to do so, and therefore, the machines do not require "sociality or respect in order to work well," [120] that they don't *care* either way. Even in designing alignment, we do have to be intentional because they "won't want to," or won't feel the desire to align – simply to be programmed to do so [120]. However, the same article (ibid.) disagrees with Omohundro's Basic AI Drives et al. by emphasizing that the AI needs to *care* to actually have goals, which would involve striving, seeking, and so on. This point seems misplaced, possibly a mincing of words, as an AI does not need to care (in the common usage of the term) in order to have goals and execute them, clearly, and the *drive* could and does come from optimization methods and achieving the goal efficiently, not *caring* in the same way that we do. One emphasis of risk here that is unique, worth mentioning, and a subtext of other fields, is the lengthy noting of possibly negative psychological issues onto humans from the effects of AI (including job replacement, integration into society, and care-taking), along with other indirect affect (ibid.).

"[T]here is no reason to think that an extremely cognitively capable agent will be informed about everything relevant to its own self-preservation" [20]. A criticism of Omohundro's analysis on AI drives argues that these drives do not apply to any design and motivation system, and that the simplest solution is to avoid autonomy in the first place, and animal-like goals. Further, solutions are provided such as adding universal constraints and closed-ended, selfless motivations [68]. Unfortunately, this still glances over Omohundro's examples of machines evading universal constraints [84] and an AI being able to cause harm with clear, selfless 'motivations'. In response to the idea that a machine's attempts to stop its termination may be unsuccessful, or unlikely, and that we are likely more capable in counteracting such a system (in that it cannot be *so much more* powerful via growth) [20], autonomy may be emergent (and incomprehensibly different to our expectations) and it is best to emphasize that any design has the potential for harm. It is best to prepare for these uncertain outcomes,[86] even outcomes where an AI is not 'aware' of why it is acting in self-preservation.

The criticism of Dr. Yampolskiy's unintended harmful robot directed at happiness is skeptical of pre-established goals attained in unpredictable ways, and doubts its level of potential of danger (by lacking actuators, relevant instruction in the program, or otherwise having ability to be harmful) – "[s]uch instruction must have been written in the robot's memory…a command must be present in their software and the command must be connected to embodied actuators" [76]. The argument in this text however fails to understand the (ultimately, effects and) creation of preservation of an ultimate goal, and the still-evident risks of an AI that do not have a physical

---

[85] "Such machines are not moral or immoral any more than a falling rock is moral or immoral." 119.     Coffee, I., *The most notable luminaries of our time are wrong to fear AI*, in *Medium*. 2016: Input Coffee. Perhaps the author is only correct when they mention that no one truly knows what will happen.

[86] This source does acknowledge that "This is not to say that one should not worry about small risks of terrible outcomes," but requests "to get a clear view of the probabilities if we are to make a qualified assessment of the expected value of working on these risks." 20.     Vinding, M. *Why Altruists Should Perhaps Not Prioritize Artificial Intelligence: A Lengthy Critique*. 2018; Available from: https://magnusvinding.com/2018/09/18/why-altruists-should-perhaps-not-prioritize-artificial-intelligence-a-lengthy-critique/.

body – of which there are numerous examples. Again, the way that an AI solves a problem or arrives at a decision is a current subject of concern due to its difficulty to understand/fully grasp and the associated risks in not truly knowing. It's the archetypal problem of the genie in the lamp (and other fabled examples), renounced [43].

Intelligence in the future is poised as open-ended, possibly existing with no goals at all. This subject often raises the question as to whether human concepts will characterize future machines, and even further, a rise to a new type of mind-space machine coming about. This specific claim is under the umbrella of intelligence as a process of ongoing growth, that development can only be in such a nonfixed way [121]. It is indeed likely that machines and their intelligences will organically evolve with development, but one *without any* goals is not on the top of the list of types to worry about or put resources towards preparation for. Even a benign AI comes along with its own set of risks. Human concepts are being consciously and unconsciously coded into current systems due to our natural bias, a phenomenon not to be ignored. A new type of machine or mind space is one of the most concerning due to its characteristic of being highly uncanny when compared with more familiar human minds and behaviors (which *still* have their clear dangers). Often a comparison to human behavior raises risk, irrespective of whether the entity is closer or further away in familiarity.

## *Gatekeeping, Ad Hominem, Authority Bias*

*This section simply highlights skeptical arguments claiming that there are only a select few individuals or groups qualified to make judgements on AI and AI risk. As with other examples of authority bias, this has no clear basis, and – even to follow the argument – some of the most 'qualified' individuals are the most concerned about these issues. We put forth constantly the position that the more educated one is in this subject, the more that they will realize that this is a topic to consider with serious weight and deep consideration.*

An argument frequently made is that since many top AI Safety researchers do not write code, they are unqualified to judge AI Risk or its correlates, citing that the claims are not based on empirical work or are unscientific [93]. This ignores well-documented existing risk in the field. However, as well, one does not need to write code in order to understand the inherent risk of AGI, just like someone doesn't have to work in a wet lab to understand the dangers of pandemics from biological weapons. This can be especially emphasized when looking at resultant effects of risk from AI (easily understood by a non-coder) and the fact that AI coders have significant blind spots that have caused risk to actualize in the field – strengthening the point that an inclusive approach when making considerations of AI and risk can be a great advantage.

Yaan LeCun – "Some people have asked what would prevent a hypothetical super-intelligent autonomous benevolent A.I. to 'reprogram' itself and remove its built-in safeguards against getting rid of humans. Most of these people are not themselves A.I. researchers, or even computer scientists" [106]. Others say that those who raise concerns are "ignorant of the realities of AI" [16]. This is countered easily by mentions of multiple AI leaders raising concerns about the realities of AI risk, and doubt in AI's capabilities is readily disproven in (current) innovative stages.

"Rather than hype fear, this is a great opportunity [to shape our ethics, morality, and ambition, all of which demand scrutiny]" [43]. It is true, that our ethical code requires an audit, but we should not pretend that there is no risk in this venture. Stuart Russell – "Senior AI researchers express noticeably more optimism about the field's prospects than was the case even a few years ago, and correspondingly greater concern about the potential risks" [43].

There is a repeated argument that the opinions that matter about gauging AI Risk are from "the computer scientists and engineers who spend their days building the smart solutions, applying them to new products, and testing them" [56]. This is often hard to parse [24] and an example of *epistemic trespassing* [122], where domain transfer is often assumed when answering hybridized questions or questions in other domains that often require both evidence and skills in two or more fields. This then leads to unjustified overconfidence (ie. the Dunning-Kruger effect) and unqualified assumptions about the subject. The expert in their own domain assumes that their evidence or acquired skills are sufficient in answering questions about the other domain or hybridized subject. It can also prove harmful in persuading non-experts or bystanders, or in wasting resources of experts needing to make repeated refutations. Computer scientists and engineers are not automatically AI Safety experts. Subjects and even sub-domains however, can appear 'near,' such as AI research generally and/or AI Safety, and yet the skill transfer is inappropriate and fails to answer questions properly or holistically. It is suggested that for example, an AI Researcher, when engaging in a cross-field effort and making an assertion about AI Safety, the judgment made to be measured by experts in the field and the original researcher's confidence in the judgement increase or decrease based on this evaluation and the skills and evidence that AI Safety Experts have (ibid.).[87]

"'Do not be fearful of AI - marvel at the persistence and skill of those human specialists who are dedicating their lives to help create it. And appreciate that AI is helping to improve our lives every day.' He [Peter Bentley] is simply offended! He and his colleagues work so hard on AI, they just want to make the world a better place, and along comes a bunch of other people who have the insolence to come and talk about AI risks. How dare they! Well, I've got news for Bentley: The future development of AI comes with big risks, and to see that we do not even need to invoke the kind of superintelligence breakthrough that is the topic of the present discussion. There are plenty of more down-to-earth reasons to be *'fearful'* of what may come out of AI. One such example…is the development of AI technology for autonomous weapons, and how to keep this technology away from the hands of terrorists" [66].

Proposals such as Draconian global surveillance can backfire as they may be perceived as a way for governments et. al to use beneficial AI to gain control. [85] Thus, careful proposals and extended research is necessary for proper AI Safety protocols and rather more importantly, trust, to be as universally beneficial as possible in the field.

In regards to public displays of AI fear lacking a concrete platform of education and discussion, and declared as being unfair in being done in such a way, arguments made that raising AI risk publicly is news manipulation and a method to win the technical arms race – such as Elon Musk who, regardless of his fears, still develops AI [123] – or increase sales (again targeting especially Elon Musk) [124] [44] [32] simply ignore the true dynamics of the field[88]. The race is currently quite hard to stop commercially, and voicing concerns are important. Education and open discussion are undoubtable necessities to the field. One note to make here is that Musk donated $10 million to keep AI beneficial [125] two years before this first argument was originally made, so the conspiracy ultimately falls flat.

---

[87] The paper mentions for the fields of say, astrology, there may well be experts however the subject itself and its validity of discovery is disputed – though the ability to recognize degeneracy is difficult, and it is often easy to simply dismiss a subject entirely. 122.   Ballantyne, N., *Epistemic Trespassing.* Mind, 2018. **128**(510): p. 367-395.

"Unfortunately, their [AI Safety] proposals constitute a non-solution to a non-problem[89]" [68]. There is the accusation that MIRI is an organization pretending to save the world, via "publishing comical pseudo—intellectual nonsense and sophistry" [126] in order to make money, with ridiculous solutions such as AI managed world dictatorships, inspiring other "half-scammy organizations" (ibid.) such as FHI [Future of Humanity Institute] and FLI [Future of Life Institute]. There is also criticism of leading minds such as Nick Bostrom fantasizing about techno-creationist views, part of a new-aged cult, and that all of the aforementioned parties do not understand transhumanism (ibid.), and simply are neo-Luddites attempting to create a new religion to "preserve the establishment" [68] and maintain the current order (ibid.). Some of the reasoning from this author derives specifically from the idea that comparing nuclear war to AI danger is ridiculous, and if true, should cause international bans on AI technology, which would make technological immortality impossible, and therefore rapid technological advances impossible (ibid.). The preferred method for addressing this form of skepticism and conspiracy, is education – especially on the real, directed risks in the field. It seems odd that organizations which are mostly non-traditional in terms of revenue (think tanks, non-profits, research institutions) would find the best way to mine dollars from the public is to pick an extremely niche, sometimes unintuitive subject to focus on and research deeply for many years, largely one that is out of the scope of common discourse.

> Unfortunately, beneficial AI is accused as being in three parts (though these are more empty attacks on character than substantial arguments) – "[a] comic book super-villain level plot to take over the world by the financial elite of the USA…;A totalitarian solution to an extremely improbable, practically non-existent risk; Luddism: preventing AI technology from being developed by anyone they do not like" [68].

Further, a common misunderstanding of transhumanism as a solution to encounters with more advanced beings – "big fish eat little fish" [80] – and that there will be dependence on humans in the equation (harm to humans is harm to the system or total relationship) [100] is that the lower, less capable being eventually is edged out in utility over time in the relationship. This can make us essentially useless and vulnerable to disposal.

## *Using History to Predict the Unpredictable Future*

---

[88] –and that other avenues are likely more marketable, such as Mars colonization. 32.    Dowd, M., *Elon Musk's Billion-Dollar Crusade to Stop the A.I. Apocalypse*, in *Vanity Fair*. 2017.

[89] Other accusations – a naïve national security mindset distorts AI risk groups' views, trying to teach human goals to an AI agent is a ridiculous contrivance, these groups are a cult of irrationality and AI eschatology, a disgrace to rationalism and consequentialism, and may have done damage to naturalist ethics, mistaking science fiction movies with reality. 68.   Özkural, E., *Scratch Artificial, Is Intelligence an Existential Risk?*, in *Medium*. 2017.

*In this section, there are repeated skeptical arguments made with the basis that the past can predict the future, that patterns of innovation that we have experienced will repeat with AI, and that previous approaches in several sectors will work in solving our future problems. This is unfounded largely for a particular reason – we are in a clearly unprecedented, accelerated, and in some cases exponentially improving time in technological history. Things are moving faster than we can adjust to, not to mention that new considerations emerge that were impossible to prepare for imaginatively (video evidence in court accused to be an AI-powered falsification being one repeated example). Problems existed in the past with technology, and as the logic follows, with more capable technologies, bigger problems may and likely will occur, with unpredictability remaining a large factor. It is unwise to to double down on avoiding risk. Of course, we as humanity should learn from the past – especially lessons from nuclear war and other catastrophic risk categories – and with that perspective in place one may would like to ask whether the below arguments by skeptics are the best possible preparations for the future.*

The analogy of AI's existential risk to nuclear war risk is flagged as unfounded, that there cannot be a mutually assured destruction scenario, "because with nuclear war, you need the threat of being able to hurt the other side whereas with AI x-risk, if the destruction happens, that affects you too" [2]. Unfortunately, the threat of extinction and mutual damage certainly *does* exist consequentially with nuclear weapons, and some advanced AI (such as autonomous weapons) do have the threat of affecting the receptive party without affecting the delivering party. The potentialities posed across this argument are all within the range of outcomes.

One claim – "[w]e're a long way off from being able to create a single human-equivalent A.I., let alone billions of them…the ongoing technological explosion will be driven by humans [(emphasizing humanlike or based decision-making)] using previously invented tools to invent new ones" [62]. This is incorrect. Developments in AI are seeing exponential growth in application and use, and more unique breakthroughs will cause and are causing more and more exponential and unprecedented progress. The unique potential for AI's explosion and spread is not to be ignored – certainly not to be underestimated.

An argument is presented by Peter Bentley that AI can only develop slowly, (or incrementally,[90] [52]) with difficulty, and will only ever grow if we want it to or force it to, (especially the *way* we want it to) paralleling it to existing intelligence and its features [56][91]. The future will be equivalent to today – as with "crops, cattle, and pets" [56] (or horses [8]), AI systems are controllable, will be manipulated to fit our needs, and be so integrated in this way that "[i]t would no more wish to kill us than it would kill itself" [56]. Along with having many assumptions, this claim is a false future equivalence, that new intelligence and its growth will be the same as current and past examples. We cannot confidently predict that the same outcome will be the case for a larger artificial 'brain,' with even less confidence in a prediction regarding AGI that is alternative from natural examples. This dilution of concern and assumption of equivalence in behavior, growth, and alignment can only create surprises, and of course, numerous counterexamples from modern (and historical) AI systems provide a counterargument by existence. Coding often includes mistakes, and discrepancies between code and our intentions (especially with goals [59], the alignment problem, and other important considerations) should be identified for their consequences and evidence as a possible lack of controlling larger and more capable systems with more sophisticated functions. Even if the equivalency were correct, there are pitfalls that already exist with current intelligence if replicated exactly into the future that can cause danger and would most definitely require safety mechanisms.

---

[90] Steven Pinker's words, adding that "[a]s AI expert Stuart Russell puts it: 'No one in civil engineering talks about 'building bridges that don't fall down.' They just call it 'building bridges.' Likewise, he notes, AI that is beneficial rather than dangerous is simply AI," 52.    Pinker, S., *We're told to fear robots. But why do we think they'll turn on us?*, in *Popular Science*. 2018. again, not understating that the subtext in this very quote is to design AI intentionally beneficial, requiring a dedicated effort to AI Safety which may be vulnerable to exponential growth and progress.

[91] "There will be no runaway AIs, there will be no self-developing AIs out of our control. There will be no singularities. AI will only be as intelligent as we encourage (or force) it to be, under duress." 56.    Bentley, P.J., et al., *Should we fear artificial intelligence?* 2018. p. 25.

This argument asks us to reflect on the lessons of history, that we should focus on new technology's effects rather than the fears and 'hysteria' surrounding it, as all inventions create good and bad results. The claim is made that regulation has historically resolved the conflict that exists with dual-use technology and it's effects, emphasizing the need for pervasive oversight at various levels, and to address the rapidity of advancement that this technology brings [127]. Though points of dual-use and addressing rapid advancement are valid, this point ignores that such a technology as AI or AGI is unprecedented in its growth, especially currently, without a historical reference. It should be intuitive then, that something as slow as regulation cannot be the full solution to a problem we can hardly keep up with. The idea that regulation has historically been a complete solution to resolve issues in technology is shortsighted. Instead, as is historically present as well with other existential risks, cooperation and coordination among the community may avoid risk and help ameliorate the aforementioned effects[92] and their own resultant effects.

One idea is that we do not need to worry about AI Safety because government regulation will intervene and prevent problems, especially since other technologies have been regulated [29]. Ethical considerations of AI are also downplayed in favor of strengthening the rule of law [128]. Given how poorly legislation against hacking, computer viruses or even spam has performed, it seems unreasonable to rely on such measures for prevention of AI risk. Regulation also has some possible pitfalls such as poor design and rapid obsolescence – and is not the only guarantee to safety, as such extrinsic measures are not enough to be effective. There are many weak points – they may be too weak or too strong due to the difficulty in assessing AI risk [58], therefore constraints need to be carefully designed, otherwise they can backfire and cause more harm[93]. Regulations may become stronger/misguided, and AI developers may become alienated and find ways to be less observable. Thus, a more mature relationship between AI experts and policy makers is needed [85] [12],[94] as well as a way for regulations and governments to keep up with ever-evolving technology, to avoid the risk of falling behind [12]. Verification of compliance and significant overhead is also solved more easily with more intrinsic motivations. Intrinsic factors and motivation also seem to be influential on the success and outcome of extrinsic measures, and ideally AI designers will choose (with genuine desire) to build the most beneficial AI design – "[i]t is difficult to overstate the importance of social norms for beneficial AI[95]" [85]. Extrinsic measures should aim to be desirable to AI developers – rules that they would want or find easy to comply with. Therefore, risk can be mitigated by better understanding AI developers. AI review boards also have an existing precedent in adjacent fields, and with review of harms, societal impacts, and ethical assumptions, such an internal mechanism could be more effective when working in tandem with regulation. With a compelling reason, such as global existential risk, self-policing and regulation[96] [58] may provide high efficacy. As this may indeed be a collective action problem[97], there are strategies beyond regulation that are of more use – including regularly correcting misconceptions that may be harmful as well as incentivizing adherence to good standards – cooperation overall being cited as a stronger incentive for an agent to reduce AI risk and create additional (expected) value,[98] as well as being able to identify appropriate norms around research [58]. There are many suggestions such as workshops, conferences, and other methods to collaborate and expand groups of beneficial AI supporters which may increase positive collaboration and make real progress in actually reducing AI risk [12, 58, 85].

---

[92] Some of the effects include job displacement, how individuals participate in the future with these technologies, and the effects on the economy that AI may provide in its exponential future. Unsafe AI would take precedent over these other problems (which may be aided by regulation, but effective regulation is preceded by cooperation).

[93] There is a running fear that regulation could cause AI companies to take flight to other countries. 58.     Amanda Askell, M.B., Gillian Hadfield, *The Role of Cooperation in Responsible AI Development.* 2019.

[94] "Create demand for people with AI background working in policy positions. Such in-house AI expertise can help ensure that AI policy matches actual issues associated with AI technology and does not inadvertently restrict beneficial AI or enable harmful AI." 12.     Baum, S.D., *Reconciliation Between Factions Focused on Near-Term and Long-Term Artificial Intelligence.* AI & Society, 2017.

"The example I [Eric Schmidt] would offer is, would you not invent the telephone because of the possible misuse of the telephone by evil people? No, you would build the telephone and you would try to find a way to police the misuse of the telephone" [129]. This statement is far from helpful. Most people are excited about new technologies and their pros, yet being in favor of benefits does not exclude or override a need for the proper building of AI, as misuse of technology is not the only risk, nor is policing the only (or most effective) solution. Even if we were aware of what the appropriate precautions exactly were, it is hard to motivate programmers in such a direction and ensure that they will follow through [49] (hardline regulatory initiatives again likely not being a complete global solution unless there is unanimous consent).

The world has never ended before, and therefore a skeptical claim is that this should apparently give us confidence about the future – not to mention that technology had been used for good and evil throughout history, but the benefits always win out by a large margin [7]. Unfortunately, unprecedented technological progress cannot be compared to lower-impact, adjacent macro-societal changes in the past, though it would be an easier viewpoint to have (however unrealistic). "[I]t's a whole new era that dawns at that point, and it is hard to foretell in much detail what it might contain" [7]. We can also observe eras of innovation that have caused – also to this day – negative effects, which, if we are wise enough, would be a pattern – if recognized – to avert.

---

[95] One example being that when constraints work, they ensure beneficial AI designs. When this happens, it limits freedom of designers, which can cause pushback, and constraints assume that AI designers do not want to choose beneficial designs. Incentives, though an extrinsic alternative, may be effective – even beyond monetary compensation (ex. social upholding which may be stronger long-term) and may reduce regulatory backlash or retaliation from researchers, however in the text incentives are described as having its own pitfalls and not the best guarantee for developing beneficial AI. 85.   Baum, S.D., *On the promotion of safe and socially beneficial artificial intelligence.* AI & SOCIETY, 2017. **32**(4): p. 543-551.

[96] Historical examples given including Responsible Care, a self-regulation program in the chemicals industry in the US. 58.  Amanda Askell, M.B., Gillian Hadfield, *The Role of Cooperation in Reponsible AI Development.* 2019.

[97] One note here – misaligned perception of risks could worsen a collective action problem, along with the race to the bottom. Building trust can, of course, improve a collective action problem.

[98] Five factors listed for detail – High Trust, Shared Upside, Low Exposure, Low Advantage, Shared Downside. 58.
       Amanda Askell, M.B., Gillian Hadfield, *The Role of Cooperation in Reponsible AI Development.* 2019.

## *Miscellaneous*

### *[Other Perspectives, Poorly-Formed Arguments, Clueless]*

There is a possible filtering mechanism regarding the topic of AI safety/risk (and the epistemology of the subject) that "deeper arguments [are] only ever exposed to people who were already, coincidentally, persuaded by the initial arguments," wherein "[b]elievers will then be in the extremely secure-feeling position of knowing not only that most people who engage with the arguments are believers, but even that, for any particular skeptic, her particular reason for skepticism seems false to almost everyone who knows its counterargument" [130]. This may give strength in addressing popular counterarguments, but many can drop out at some point in the argument chain, even if it is not eventually logical to do so (ibid.). This could be a case as to why many of these skeptical arguments, which are exceptionally poor in strength, are displayed out in the public with some authority even at present.

"In his eagerness to establish that a dangerous AI breakthrough is unlikely and therefore not worth taking seriously, Michael Shermer holds forth that work on AI safety is underway and will save us if the need should arise…overlooking that it is *because such AI risk is taken seriously* that such work comes about" [107].

Andrew Ng remarks that computers are becoming more intelligent, "[b]ut sentience and consciousness is not something that most of the people I talk to think we're on the path to" [106]. Accidental emergence aside, non-conscious intelligence (and non-conscientious development) still poses great risk. It only takes one clever person or group to have this drive and intention for it to be flagged and addressed.

Complex minds will likely have complex motivations, and this may well be what it means to be intelligent[99]. This statement may not be a direct skepticism of AI risk, but is skeptical of the ability of AI Safety to resolve this issue. Complex goals also don't imply that they are humane goals [34][100].

One argument is that certain narratives about AI (including the historical fact that the term itself can be controversial and muddling [21]) can influence AI researchers, and "influence public acceptance and uptake of AI systems," being in favor of a more positive take on AI (and its benefits) displayed to the public[101] [65] – "[r]egulation [due to AI or AGI fear] can discourage innovation" [58] or stifle progress[102] [102] – "regulation stifles innovation…AI becomes a dirty phrase…[a]nother AI Winter [resulting from overpromising and disappointment [43]] destroys progress…[d]o not be fearful of AI[103]" [56][104]. Though a tangential point, there is also criticism of current recommender systems[105] [43] and other examples accused as being propped up as more sophisticated than they in actuality are. The strategy and mindset behind these references will likely backfire. When an overwhelmingly positive take is given to the public, such as AI, with the benefits and possibilities displayed at the forefront, and an eventual issue does occur which harms humans, it will appear catastrophic and likely reduce trust in both the field across sectors – unreasonably so – and damage the reputation of the respective companies [58]. It is also potentially more difficult for the public to themselves assess AI Safety than as with other sectors [58]. Hiding potential issues and avoiding regulation and AI Safety mechanisms will actually reduce funding and merit a worse public opinion long-term. In addition, avoiding addressing risk historically has not been optimal for funding [16]. It is possible that regulating extremely high-risk AI systems, as proposed in the Artificial Intelligence Act, may create more public trust in oft-used, low-risk AI systems and the overall field. It is also possible that, as with the events of the 1975 Asilomar conference regarding recombinant DNA, successful coordination could come directly from "substantial buy-in from the relevant scientific community,[106]" [10], more informal social mechanisms (ibid.), incentives [58], and consensus within communities of people [131] (notably AI communities) [59] [48] [12] rather than solely via strict oversight[107] [48]. Interestingly enough, a wider audience engagement may resolve many issues regarding pertinence, safety, and blind spots in risk[108] [40]. Thus transparency within reason[109] is a better emphasized method and may also reduce misunderstandings and unintentional defection (both of which increase risk) [58].[110]

---

[99] According to the source, this appears in AI alignment in two places – where simple proposals of what an AI should care about will not work, because human values/motivations are complex, and that AI will likely have convergent instrumental goals, finding the common strategies to complete such goals. 34. Graves, M. *Response to Cegłowski on superintelligence*. MIRI 2017; Available from: https://intelligence.org/2017/01/13/response-to-ceglowski-on-superintelligence/.

[100] "For example: Suppose an agent is considering two plans, one of which involves writing poetry and the other of which involves building a paperclip factory, and it evaluates them based on expected number of paperclips produced (instead of whatever complicated things motivate humans). Then we should expect it to prefer the second plan, even if a human can construct an elaborate verbal argument for why the first is 'better.'" 34.    ibid.

---

[101] Includes the fear "that developments in AI might be threatened with the kind of public hostility directed towards genetically modified (GM) crops. This would also apply to AI system regulation, as they claim narratives would shape views of policymakers and constituents." 65. Cave, S. and K. Dihal, *Hopes and fears for intelligent machines in fiction and reality.* Nature Machine Intelligence, 2019. **1**(2): p. 74-78.

[102] One note here is that Elon Musk and others' raised concerns could be accelerating AI research by increasing interest in the field 32. Dowd, M., *Elon Musk's Billion-Dollar Crusade to Stop the A.I. Apocalypse*, in *Vanity Fair*. 2017., which, with proper safety mechanisms in place, would be a benefit.

[103] Similar issues have received substantial attention (notably past global risks) 48. Seth D. Baum, R.d.N., Anthony M. Barrett, Gary Ackerman, *Lessons for Artificial Intelligence from Other Global Risks.* The Global Politics of Artificial Intelligence: p. 21.

[104] It does mention here that "we should be focussing [sic] on new safety regulation and certification for each specific safety-critical application of AI," as well as identifying a risk that humans may be less competent when the AI requires a task to be taken over by humans (e.g. driverless cars). 56. Bentley, P.J., et al., *Should we fear artificial intelligence?* 2018. p. 25. This may be identified as embedded skepticism or a gray area.

[105] Introduces the debate of measurement vs. manipulation. 43. Lanier, J., *The Myth of AI*, J. Brockman, Editor. 2014: Edge.

[106] "–those who worry about AI impacts should not drain their energy on internal disputes."12. Baum, S.D., *Reconciliation Between Factions Focused on Near-Term and Long-Term Artificial Intelligence.* AI & Society, 2017.

[107] –Or a combination of both. References include the lasting effects over several decades regarding GMO regulations. Building an effective path dependency may create fewer problems in the future, or provide fertile ground for policy. 132. Nicolas Moës, J.B., *AI Alignment Podcast*, in *On the Long-term Importance of Current AI Policy with Nicolas Moës and Jared Brown*, L. Perry, Editor. 2020. This may also fit in with AI Regulation Will Solve Problems.

[108] This criticism in part references the concerns raised in the historical biotechnology development discussions and looks at the AI Principles as being limited in having diverse perspective – encouraging democratic engagement, (however uncomfortable it may be). 40. Jack Stilgoe, A.M., *It's time for some messy, democratic discussions about the future of AI*, in *The Guardian*. 2017.

One Steven Pinker is afraid of making others aware of the risk of dual-use technologies, as it may increase the likelihood of danger – "[s]owing fear about hypothetical disasters, far from safeguarding the future of humanity, can endanger it" [46]. One may fear that sounding the alarm could create chaos. It matters more *how* we warn the public and that educating the public can be empowering rather than a negative effect, and that public statements about existential risk have been more clinical and of more merit than fearmongering or hyperbole – as such concerns have historically predicted or prevented disaster, leaving the question of any validity of the above claim (ibid.).

Do not even talk about superintelligence, skeptics claim – some current researchers can resent even the mention of long-term AI, with claims of fear mongering [93], fears of slow adoption [102] or another cycle of hype and disillusionment [12] [8] – that "the hype is the reason why AI research dives into periods of recession" [56]. John Giannandrea calls hype around AI and AGI "unwarranted and borderline irresponsible" [39]. One author even suggests to encourage AI/ML experts to downplay AGI in order to calm the public [102], ignoring the fact that experts in the field have already deemed it a great risk needing attention, not avoidance. Ignoring an existential threat (or even AI's capabilities) is, again, a risky choice, and it is hard to argue that the right solution is to have a complete diversion of focus. Such a method may also tank research efforts and long-term funding. Pinker claims that discussing another global risk will overwhelm people and cause them to lose faith in the future. Häggström responds[111] that only the threat being real is enough to justify the need for an open discussion, as ignoring the threat will again be a way for the small chance of more serious, global risk occurring to actually happen or have higher probability of happening (notwithstanding technological benefits being attractive). Even if we were still theoretically unsure of the risk, an open discussion would *still* remain important to address its weight from experts [56]. By contrast, with AI systems in general, openness can allow for (or boost) the field's reputation to stabilize cooperative efforts [58] and potentially increase funding. Another point here, as seen historically with concerns regarding asteroid collision, is that the risk being rare does not exclude it from being taken seriously, especially in regards to policymaking [48]. With the subject of superintelligence and AI safety, the stakes of the game could hardly be higher [16] [133] and the risks are worth mentioning – and addressing [46] – above fear of judgement.

A message given to Congress – that the "real disaster would be abandoning or inhibiting cognitive technology before its full potential can be realized" – calling the umbrella of issues (long- and short-term) "fantas[ies]" [53]. Such a claim ignores that without safety mechanisms in place, 'full potential' will simply *never be realized*.

---

[109] Releasing private data and AI systems vulnerable to misuse are parts of transparency that may cause risk. 58.   Amanda Askell, M.B., Gillian Hadfield, *The Role of Cooperation in Reponsible AI Development.* 2019.
[110] One proposition – "if future AI programs learn of rival programs' struggles and cessations, then they may also stop or proceed more carefully. This possibility runs counter to the proposed idea that information about rival AI projects increases risks." 58.   ibid.
[111] Häggström also mentions that such a line of thinking counters Pinker's own beliefs of intellectual openness. 56.   Bentley, P.J., et al., *Should we fear artificial intelligence?* 2018. p. 25.

The degree to which an idea 'sounds crazy' is not a criterion for its illegitimacy. Some critics will dismiss an idea about AI risk that they don't understand simply because it sounds too far out, futuristic, or even claim that things are getting better (without providing evidence). [46] However, one perspective is that "[i]f you were writing during the early days of the scientific method, or at the dawn of the Industrial Revolution, then an accurate model of the world would require you to make at least a few extreme-sounding predictions" [34].

There are numerous arguments claiming that the problem of AI Risk, and related legislation – especially when dealing with existential AGI risk – is full of "non-existent problems" and "has no basis in reality, but is mere science fiction" [56]. Jaron Lanier says that beliefs in development of Superintelligence are expressions of a new religion "through an engineering culture" [61]. The best countermeasure for such denialism without basis, is, again, education (especially via cited sources and historical examples), characterization, raising awareness [48], and correcting false beliefs [58]. "We cannot assume AI to be a discipline that is only beneficial to humanity" [1]. Education may also create a taboo against misuse between countries and prevent a (destructive) race from occurring [134].

There is a statistical underestimation among researchers of catastrophic risk regarding and lack of priority of research into AI safety [135]. Societal significance of AI research may be considered less important to researchers, and because it may be outside of their professional domain, they may not even form opinions and/or dismiss the subject, prioritizing AI's capabilities. When they do focus on these issues, it is usually on near-term problems as they will be more commonly accepted – "AI researchers, like other scientists and engineers or any other people, are not above the law or above morality. It is everyone's responsibility to help society," [12] regardless of whether the focus is on current innovation or future impacts, and the best opportunities for that future will likely arise with alignment (ibid.) in the field.

Overambitious tasks given to Superintelligent systems may go against the goal of improving human and/or sentient well-being, and likely starting with modest, non-open ended goals may be "less hazardous to get wrong" [34].

On the opposite side of the argument regarding regulations, is also argued that regulations are both unnecessary and harmful, as regulations shackle innovation, slow income growth, and keep workers out of the labour market – "Techno-Jeremiahs…populist, neo-Luddite ("Luddites"[112] including Elon Musk, Stuart Russell, Bill Gates – even Alan Turing[113] [16]) backlash [is] against the next production revolution…[we should] *accept the hypothetical risk posed by technology*. Imposing restrictive regulations on technologies based on speculative fears would only slow their development and limit their benefits. Countries should instead embrace the innovation principle, which says that policymakers should address risks as they arise, or allow market forces to address them, and not hold back progress because of hypothetical concerns." [50]. Aside from sweeping under the carpet well-documented *existing* AI risks, this argument is likely related to one geographical territory or country falling behind another in terms of innovation and technological progress, and this is especially true with any marketable AI in the past, present, and future. For several reasons when considering the field long-term, it is possible that a first-mover advantage may not even exist within AI development. AI companies do have incentives to develop systems responsibly, but they are weaker incentives (at first glance) than in other industries, and competitive pressure compounds this weakness [58]. AI progress being non-linear in growth (discontinuous) means that ex-post measures such as market response and liability will be less effective [58] or reliable. The issue with a winner-takes-all technology race is that it makes a "supposedly dangerous technology seem desirable" [85], and the rush to win may be fatal before any benefits are realized,[114] real existential or otherwise risk being greater than the risk of being in second place.

---

[112] –even claiming that the entire field of AI existential risk or "AI eschatology" is a case of luddite cognitive bias.
93.        Özkural, E., *Censored Criticism on the Wikipedia Article About AI Eschatology*, in *log.examachine.net*. 2017.
[113] "It is as if one were to accuse nuclear engineers of Luddism if they point out the need for control of the fission reaction." 16.        Russell, S., *Provably Beneficial Artificial Intelligence*.

Building AI systems simply through the lens of technological innovation (or even for its own sake or for intellectual merit [12]) is a shortsighted drive, and yet building ever more capable systems remains the current focus of the field. This shelves or ignores safety mechanisms and their related implications. This 'race to the bottom,' especially when one is aware of AI Safety, is dangerous and has reduced net positive gain for both companies and consumers, where defecting from cooperation will only produce more risky systems without an upside – yet consequently, the leap to cooperate can cause more benefits to all actors [58]. For example, potential benefits that exist with nuclear power have not been realized (and the field's reputation damaged) due to the lack of attention or care to its risks *proactively*. Objectors to safety concerns may call developers "anti-AI," akin to calling nuclear engineers "anti-physics" [16], and the potential success of AI (which is massive) is dependent on the sufficient addressment and prevention of its potential risks (ibid.). The highlighted risks of incautious development cannot be ignored [58]. As Stuart Russell mentions, beneficial AI needs to be baked into the field of AI itself and create a new standard and social norm for the field – for example, in civil engineering, no one is centering the discussion and debate around "building bridges that don't fall down" [85] – they are simply calling their work "building bridges," and this would normalize researchers seeking to benefit society through such work (ibid.).

Businesses "may wish to demonstrate that they are responsible actors on AI and therefore do not need to be regulated,[115]" [48] and such pushback may be due to the start of regulation via ethical considerations (ibid.). This motivation fails to understand that market forces and corporations are not the best litmus tests of risks of AI, and that the race itself to get to market can be self-destructive. A market where AI is deployed as quickly as possible without proper safety measures will quickly reduce the level of quality of these products, cause significant harm, and reduce consumer trust (not to mention the personal-to-existential risk) – causing definite backfire. Improving expert consensus on AI risk may prevent corporations from obfuscating its validity, weight, and current significance, as was done with the risk of climate change [48]. It is also clear that addressing risks as they arise will get out of hand very quickly and some effects may be hard to reverse, exemplified by data manipulation, privacy loss, and/or physical accidents. Concerns about AI often deal with accidental harm [48], and thus being proactive and timely is more of a priority than with other sectors. For example, regarding autonomous vehicles, "[a]utomobiles must be safe, otherwise they will not sell and manufacturers can face steep liability claims…'You can't hit control-alt-delete when you're going 70 miles an hour' [Ford CEO Mark Fields]" [85].

---

[114] "AI groups may have some desire for the power that might accrue to whoever builds strong AI, but they presumably also desire to not be killed in the process." 85.    Baum, S.D., *On the promotion of safe and socially beneficial artificial intelligence.* AI & SOCIETY, 2017. **32**(4): p. 543-551.

[115] It is noted that corporations may push the narrative that they are more responsible than other companies in this regard. 48.    Seth D. Baum, R.d.N., Anthony M. Barrett, Gary Ackerman, *Lessons for Artificial Intelligence from Other Global Risks.* The Global Politics of Artificial Intelligence: p. 21.

One problem noted in government regulation is that it can cause a 'brain drain' effect, where more work moves off-shore, and prevents regulated oversight, where and while unseen development continues, therefore increasing risk [43].

Some scientists, when sounding the alarm, may be afraid of losing funds [11] for more pure scientific research as was feared with other existential risks – the "giggle factor" [48] – or even "risk losing [their] respectability" [12]. It appears that this argument is reputation-based, heavily weighting non-expert opinions, and would fall apart if such scientists realized that not taking AI risk seriously is more embarrassing and costly long-term in its results than taking it too seriously. Though pure scientific research may have an uncorruptible aim at the truth, to begin a project without considering seriously its risk in eliminating humanity (or making it worse off) is hard to defend as acceptable or permissible. Key in point example – "[a]nyone who contributes to the AI arms race of autonomous weapons is making the world worse" [59].

The collaborative alternative in this field – by institutionalizing a monopoly on superintelligence development – may have the perverse result of increasing existential risk due to a lack of competitor superintelligences that could keep their "fellows" in check [91]. One thought experiment – "[s]topping everything is not feasible short of imposing a global police state of some sort, which would be very difficult, highly ethically questionable, and probably wouldn't work anyway. Stopping just AGI or (say) just nanotech and continuing everything else would have quite dubious value, as in the overall context of advanced tech development, each individual technology has both dangers and defensive potential" [121]. Aside from extremes, it seems that a healthy level of surveillance and checks would reduce risk and increase the chances of creating a beneficial AI.

This source asserts that "[i]f a project can't innovate faster than the world, it can't grow faster to take over the world" [136]. Isolation may prevent others from learning about its progress, but may lose revenue, and it would have to be much better than the rest of the world at improving software. There is also not one specific place to improve intelligence (ibid.) As mentioned, the dismissal of exponentiality is a key reason as to why some fears are overlooked, and such intelligences can improve over a variety of factors and competence. In addition, it appears that in an interconnected world, true isolation can never exist for machines nor innovation.

One skeptical claim purports that powerful agents will appear coherent to us, because if we can find incoherency, so can they – as not all behavior is goal-directed, and most arguments start with the assumption of goal-directed behavior and derive expected-utility maximizers. Apparently , there is no reason why Superintelligence would have goals [137]. We currently build AI systems to have goals, and see little reason not to continue this trend – though it should also be repeated that a non-doer is not a Superintelligence to prioritize our preparation. Perhaps a benign AI can do secondary damage due to its immense potential as a tool or resource by a certain individual or group while not being an agent itself.

There is criticism that anticipating a threshold, and the end of the world, such as a singularity, are all recycled subtexts [43]. "It's surprising, by the Copernican principle, that our time looks as pivotal as it does. But while we should start off with a low prior on living at a pivotal time, we know that pivotal times have existed before, and we should eventually be able to believe that we are living in an important time if we see enough evidence pointing in that direction" [34]. It is a difficult situation to understand, in being unprecedented, and our intuitions are likely to break down in such a scenario [32].

"Many arguments for working on AI safety trade on definition tricks, where the sentences 'A implies B' and 'B implies C' both seem obvious, and this is used to argue for a less obvious claim 'A implies C'; but in fact 'B' is being used in two different senses in the first two sentences"[116] [34]. This is true for low-grade futurism, however less formal terms or understanding do not undermine the arguments themselves. We also have to consider that argument and theory are more relevant in problems such as AGI Safety, because testing hypotheses in this field can be much more dangerous in outcome (and process) than with other fields (ibid).

Steven Pinker aligns with the idea of an optimistic future – and the reality of our increased well-being, though he may not understand that lowered violence and similar realities (in part due to technology) are not contradictory to the highest global risk potential that we have ever faced. Additionally, a lack of precedence of the current situation should raise more caution and attention, not less, towards the issue and how risk can best be avoided [46].

There are several ways that autonomous weapons[117] can be uniquely dangerous, especially when self-contained (without a human in the loop of the decision-making process). Accidents, hacking, lack of explain-ability, lack of moral context, speed of decision making, ability of escalation [138], and numerous more may have consequences before the weapon can be reprogrammed in time.

This argument asserts that via the precautionary principle, a preservation of boundaries would prevent innovation in the field, and a proactionary approach is better. We must build AGIs which are intrinsically cooperative, rather than bounded. For example, the paperclip maximizer is criticized as a "stupid goal," and the AGI might reject it if the system were built to be more cooperative – AGIs are likely to form their own goal systems as "ongoing interactions with humans in human society[118]," and "we should just nudge in the right direction as it progresses" [121]. 'Just build it safely!' is the entire *modus operandi* of this field and these efforts, and it is possible to increase safety without destroying innovative initiatives.

---

[116] "For example, when building sand castles it's low cost to test a hypothesis; but when designing airplanes, full empirical tests are more costly, in part because there's a realistic chance that the test pilot will die in the case of sufficiently bad design. Existential risks are on an extreme end of that spectrum, so we have to rely particularly heavily on abstract argument (though of course we can still gain by testing testable predictions whenever possible)."
34.     Graves, M. *Response to Cegłowski on superintelligence*. MIRI 2017; Available from: https://intelligence.org/2017/01/13/response-to-ceglowski-on-superintelligence/.

[117] Many arguments in the field will justify development of autonomous weapons due to the fact that other regions are also producing the same weapons, and, to repeat, the arms race is an extremely risky path for humanity to pursue.

Roko's Basilisk being a case against attending to existential risk [93] is easily countered by the perspective that AI Safety is a way to redirect and not restrict progress, and therefore in this instance it would not be a condition for demise. It is also impossible to determine the interpretability of an AGI and how it will decide what exists as obstacles to its growth. Hedging humanity's survival on this bet is a foolish idea without substance – "[w]hy would a *deity*-like superintelligence be bound to an inflexible, rote, unempathetic logic that led it to alter history with dictatorial brutality, an 'if-then' equation gone feral?[119]" [81]

Skepticism towards technology is purported as simply a Western philosophy – we are used to a warlike culture and hostile view of AI – and thus, an obsession with either utopia or apocalypse is bred [139]. Though this line of thinking may be useful for a broader discussion of ethics, as Eastern philosophies may provide useful solutions and alternative perspectives, alternative perspectives do not eliminate the existing risks of artificial intelligence. Even with the above possibility, we cannot 'bet' on AI achieving Enlightenment.

## Conclusions

Unmitigated skepticism regarding the potential risks of AI systems has the potential to have a detrimental impact on the long-term flourishing of humanity. This skepticism questions the necessity of incorporating necessary safeguards into intelligent systems, and creates a misplaced doubt and reduction of attention with regards to the allocation of resources for ensuring the safety and security of such systems. It is, therefore, of the greatest importance that we take steps to explicitly call out instances of AI risk skepticism and address them head-on.

Both the development and deployment of AI represent one of the most significant technological advancements of all-time. As such, it is imperative that we approach this development with the appropriate amount of caution, and that we take necessary steps to better ensure the safety and security of the intelligent systems that we are creating. This means allocating the necessary resources, developing appropriate safeguards, and engaging in open and honest discussions about the potential risks associated with AI. The unchecked skepticism of AI risk represents a threat to the long-term flourishing of humanity. It then becomes our duty to address this issue with the seriousness and weight that it carries, and to take all necessary steps to ensure the responsible and ethical development of AI technologies.

---

[118] "[I]f one adopts a pragmatic rather than idealistic attitude to agency (i.e. looks at what can plausibly be done by real-world individuals or organizations, rather than what should ideally be done by humanity as a whole according to certain aspirations), then the conflict between conservative humanity-wide goals and more radical Cosmist goals largely disappears." 121.   Goertzel, B., *Superintelligence: Fears, Promises and Potentials | Reflections on Bostrom's Superintelligence, Yudkowsky's From AI to Zombies,
and Weaver and Veitas's "Open-Ended Intelligence"*. Journal of Evolution and Technology, 2015. **24**(2): p. 55-87.
[119] Further exploration of this logic would show that a rogue AI, *not* listening to its 'if-then' equation, is another example of danger in this situation.

# BIBLIOGRAPHY

1. Verdicchio, M., *Will we be masters or slaves of the Information Technology of the future?* 2018.
2. Asya Bergal, R.L., Sara Haxhia, Rohin Shah, *Conversation with Rohin Shah*, in *AI Impacts*. 2019.
3. AI, A., *AI & Business | A Practical Guide | Part One: An Introduction to AI | AI 101*, A. AI, Editor.
4. Hamrud, E. *AI Is Not Actually an Existential Threat to Humanity, Scientists Say*. 2021 11 April 2021; Available from: https://www.sciencealert.com/here-s-why-ai-is-not-an-existential-threat-to-humanity.
5. Gilbert, B. *Elon Musk tweets 'Facebook sucks' at Facebook AI lead who said Musk 'has no idea what he is talking about when he talks about AI'*. 2020; Available from: https://www.businessinsider.com/elon-musk-says-facebook-sucks-argument-with-ai-lead-2020-5.
6. Boden, M.A. *Mapping the Future of AI*. 2017; Available from: https://www.project-syndicate.org/commentary/artificial-intelligence-no-singularity-by-margaret-a--boden-2017-02.
7. Spectrum, I. *HUMAN-LEVEL AI IS RIGHT AROUND THE CORNER—OR HUNDREDS OF YEARS AWAY*. IEEE Spectrum 2017; Available from: https://spectrum.ieee.org/humanlevel-ai-is-right-around-the-corner-or-hundreds-of-years-away.
8. Brooks, R., *The Seven Deadly Sins of Predicting the Future of AI*, in *Rodney Brooks | Robots, AI, and other stuff*. 2017.
9. Shermer, M. *Artificial Intelligence Is Not a Threat—Yet*. 2017; Available from: https://www.scientificamerican.com/article/artificial-intelligence-is-not-a-threat-mdash-yet/.
10. Kurzweil, R., *Don't Fear Artificial Intelligence*, in *TIME*. 2014.
11. *Existential risk from artificial general intelligence*, in *Wikipedia*. Wikipedia.
12. Baum, S.D., *Reconciliation Between Factions Focused on Near-Term and Long-Term Artificial Intelligence.* AI & Society, 2017.
13. Tegmark, M. *The Top Myths About Advanced AI*. 2016; "Resource"]. Available from: https://futureoflife.org/resource/aimyths/.
14. Mitchell, M., *Why AI is Harder Than We Think.* 2021.
15. Dario Amodei, C.O., Jacob Steinhardt, Paul Christiano, John Schulman, Dan Mané, *Concrete problems in AI Safety.* 2016.
16. Russell, S., *Provably Beneficial Artificial Intelligence.*
17. Hamblin, J., *But What Would the End of Humanity Mean for Me?*, in *The Atlantic*. 2014.
18. Etzioni, O., *How to know if artificial intelligence is about to destroy civilization*, in *MIT Technology Review*. 2020.
19. Littman, M., *Interview with Michael Littman on AI risks*, A. Kruel, Editor. 2015.
20. Vinding, M. *Why Altruists Should Perhaps Not Prioritize Artificial Intelligence: A Lengthy Critique*. 2018; Available from: https://magnusvinding.com/2018/09/18/why-altruists-should-perhaps-not-prioritize-artificial-intelligence-a-lengthy-critique/.

21. Brooks, R., *The Seven Deadly Sins of AI Predictions*, in *MIT Technology Review*. 2017.
22. Dai, W. *Reframing the Problem of AI Progress*. 2012; Available from: https://www.lesswrong.com/posts/nGP4soWSbFmzemM4i/reframing-the-problem-of-ai-progress.
23. Schulman, J., *Frequent arguments about alignment*, in *Alignment Forum*. 2021.
24. Cremer, C.Z., *Deep limitations? Examining expert disagreement over deep learning.* Progress in Artificial Intelligence, 2021.
25. Dembski, W.A. *Artificial Intelligence: Unseating the Inevitability Narrative*. 2021; Available from: https://mindmatters.ai/2021/04/artificial-intelligence-unseating-the-inevitability-narrative/.
26. Lee, K.-F., *The Real Threat of Artificial Intelligence*, in *The New York Times*. 2017.
27. Kaufman, J. *Conversation with Bryce Wiedenbeck*. 2017; Available from: https://www.jefftk.com/p/conversation-with-bryce-wiedenbeck.
28. Kaufman, J. *Conversation with Michael Littman*. 2017; Available from: https://www.jefftk.com/p/conversation-with-michael-littman.
29. Clint, E., *Irrational AI-nxiety*, in *Quillette*. 2017.
30. Winfield, A., *Artificial intelligence will not turn into a Frankenstein's monster*, in *The Guardian*. 2014.
31. Walsh, T., *Elon Musk is wrong. The AI singularity won't kill us all*, in *WIRED*. 2017.
32. Dowd, M., *Elon Musk's Billion-Dollar Crusade to Stop the A.I. Apocalypse*, in *Vanity Fair*. 2017.
33. Fagella, D. *Is AI Safety Research a Waste of Time?* 2018; Creating or Enhancing Consciousness, Reflecting on What I've Read, The Good Must Be Explored]. Available from: https://danfaggella.com/is-ai-safety-research-a-waste-of-time/.
34. Graves, M. *Response to Cegłowski on superintelligence*. MIRI 2017; Available from: https://intelligence.org/2017/01/13/response-to-ceglowski-on-superintelligence/.
35. Cegłowski, M., *Superintelligence | The Idea That Eats Smart People*, in *Idle Words*. 2016.
36. Socher, R., *Commentary: Fear of an AI Apocalypse Is Distracting Us From the Real Task at Hand*, in *Fortune*. 2018.
37. Vasilaki, E. *Worried about AI taking over the world? You may be making some rather unscientific assumptions*. 2018; Available from: https://theconversation.com/worried-about-ai-taking-over-the-world-you-may-be-making-some-rather-unscientific-assumptions-103561.
38. Knight, W., *Forget Killer Robots—Bias Is the Real AI Danger*, in *MIT Technology Review*. 2017.
39. Vincent, J., *Google's AI head says super-intelligent AI scare stories are stupid*, in *The Verge*. 2017.
40. Jack Stilgoe, A.M., *It's time for some messy, democratic discussions about the future of AI*, in *The Guardian*. 2017.
41. Alexander, S., *Contra Acemoglu On...Oh God, We're Doing This Again, Aren't We?*, in *Astral Codex Ten*. 2021.
42. Özkural, E., *Is AI doomsaying bad science fiction?* log.examachine.net, 2014.
43. Lanier, J., *The Myth of AI*, J. Brockman, Editor. 2014: Edge.
44. O'Neil, C., *Know Thy Futurist*, in *Boston Review*. 2017.

45. Kaj Sotala, S.K., habryka, Daniel Trenor, Michael Anissimov. *AGI Skepticism*. Available from: https://www.lesswrong.com/tag/agi-skepticism.
46. Torres, P., *A Detailed Critique of One Section of Steven Pinker's Chapter "Existential Threats" in Enlightenment Now*, in *Project for Future Human Flourishing*.
47. Knapton, S., *Artificial Intelligence is greater concern than climate change or terrorism, says new head of British Science Association* in *The Telegraph*. 2018.
48. Seth D. Baum, R.d.N., Anthony M. Barrett, Gary Ackerman, *Lessons for Artificial Intelligence from Other Global Risks.* The Global Politics of Artificial Intelligence: p. 21.
49. Piper, K. *The case for taking AI seriously as a threat to humanity*. 2020; Available from: https://www.vox.com/future-perfect/2018/12/21/18126576/ai-artificial-intelligence-machine-learning-safety-alignment.
50. Atkinson, R., *Don't Fear AI*. Big Ideas. 2019. 37.
51. Reese, H., *IdeaFestival speaker on the myths surrounding artificial intelligence*, in *EO Weekly*. 2017.
52. Pinker, S., *We're told to fear robots. But why do we think they'll turn on us?*, in *Popular Science*. 2018.
53. Romm, T. *IBM is telling Congress not to fear the rise of an AI 'overlord'*. 2017; Available from: https://www.vox.com/2017/6/27/15875432/ibm-congress-ai-job-loss-overlord.
54. Dorr, A., *Common errors in reasoning about the future: Three informal fallacies.* Technological Forecasting and Social Change, 2017. **116**: p. 322-330.
55. Delaney, J.K., *Time to get smart on artificial intelligence*, in *The Hill*. 2017.
56. Bentley, P.J., et al., *Should we fear artificial intelligence?* 2018. p. 25.
57. Floridi, L. *Charting Our AI Future*. 2017; Available from: https://www.project-syndicate.org/commentary/human-implications-of-artificial-intelligence-by-luciano-floridi-2017-01.
58. Amanda Askell, M.B., Gillian Hadfield, *The Role of Cooperation in Reponsible AI Development.* 2019.
59. Häggström, O., *Science for good and science for bad.*
60. *Imbecilic Dog-God-AI-Delusion*, in *SingularityUtopia*. 2015.
61. Fjelland, R., *Why general artificial intelligence will not be realized.* Humanities and Social Sciences Communications, 2020. **7**(1): p. 10.
62. Chiang, T., *Why Computers Won't Make Themselves Smarter*, in *The New Yorker*. 2021.
63. Harris, S., *Can we build AI without losing control over it?*, in *TED Conferences*. 2016: YouTube.
64. Romero, A., *Here's Why We May Need to Rethink Artificial Neural Networks*, in *Medium*. 2021.
65. Cave, S. and K. Dihal, *Hopes and fears for intelligent machines in fiction and reality.* Nature Machine Intelligence, 2019. **1**(2): p. 74-78.
66. Häggström, O., *A spectacularly uneven AI report*, in *Häggström hävdar*. 2018.
67. Adriana Braga, R.K.L., *The Emperor of Strong AI Has No Clothes: Limits to Artificial Intelligence.* Information, 2017.
68. Özkural, E., *Scratch Artificial, Is Intelligence an Existential Risk?*, in *Medium*. 2017.
69. Kelly, K., *The Myth of a Superhuman AI*, in *Wired*. 2017.
70. Think, B., *Michio Kaku: How to Stop Robots From Killing Us*. 2011: YouTube.

71. Perez, C.E., *The Possibility of a Deep Learning Intelligence Explosion*, in *Medium*. 2017: Intuition Machine.
72. Sotala, K. *14 objections against AI/Friendly AI/The Singularity answered*. 2007; English Essays].
73. Chollet, F., *What worries me about AI*, in *Medium*. 2018.
74. Bilton, N., *Artificial Intelligence as a Threat*, in *The New York Times*. 2014.
75. Sofman, B., *Yes, We Will Live With Artificial Intelligence. But It Will Be Friend, Not Foe.*, in *Medium* 2015: Backchannel.
76. Johnson, D.G. and M. Verdicchio, *Reframing AI Discourse.* Minds and Machines, 2017. **27**(4): p. 575-590.
77. *The Batch | October 30, 2019*. 2019; Essential news for deep learners]. Available from: https://info.deeplearning.ai/the-batch-daemon-spawn-agi-takeover-deepfake-deluge-bias-crisis-how-scared-should-you-be?utm_source=Ngsocial&utm_medium=facebook&utm_campaign=TheBatchArchiveOctober312019&fbclid=IwAR1E8g2cc19FOuf2X1sVSQ4_ywFjfHdWl5F8nZWoV9KAUCn1WWQzKS1xZiM.
78. Lewis, J.A. *Waiting for Skynet*. 2018; Available from: https://www.csis.org/analysis/waiting-skynet.
79. Simon Johnson, J.R. *Jobs in the Age of Artificial Intelligence*. 2017; Available from: https://www.project-syndicate.org/commentary/artificial-intelligence-will-not-kill-jobs-by-simon-johnson-and-jonathan-ruane-2017-05.
80. Messerly, J.G. *'I'm Glad the Future Doesn't Need Us' - Bill Joy's Pessimistic Futurism*. 2016; Available from: https://archive.ieet.org/articles/messerly20160224.html.
81. Cross, K. *The existential paranoia fueling Elon Musk's fear of AI*. 2018; Available from: https://www.documentjournal.com/2018/04/the-existential-paranoia-fueling-elon-musks-fear-of-ai/.
82. Yarvin, C., *There is no AI risk*, in *Gray Mirror*. 2021.
83. XiXiDu. *No Basic AI Drives*. 2011; Available from: https://www.lesswrong.com/posts/RQWMDhhRPEf5AXtg3/no-basic-ai-drives.
84. Omohundro, S.M., *The Basic AI Drives.* Self-Aware Systems, Palo Alto, California.
85. Baum, S.D., *On the promotion of safe and socially beneficial artificial intelligence.* AI & SOCIETY, 2017. **32**(4): p. 543-551.
86. Chollet, F., *The implausibility of intelligence explosion*, in *Medium*. 2017.
87. Hawkins, J. *The Terminator Is Not Coming. The Future Will Thank Us.* 2015; Available from: https://www.recode.net/2015/3/2/11559576/the-terminator-is-not-coming-the-future-will-thank-us.
88. Torres, P., *Why superintelligence is a threat that should be taken seriously.* Bulletin of the Atomic Scientists, 2017.
89. TURING, A.M., *I.—COMPUTING MACHINERY AND INTELLIGENCE.* Mind, 1950. **LIX**(236): p. 433-460.
90. Garis, H.d., *THE SINGHILARITY INSTITUTE: My Falling Out With the Transhumanists*, in *h+ Magazine*. 2012.

91. Karlin, A. *The Case for Superintelligence Competition*. 2017; "An Alternative Media Selection"]. Available from: https://www.unz.com/akarlin/superintelligence-competition/.
92. Aum, S., *Please, No More Mentioning of 'AI Safety'!*, in *Medium*. 2017: Intuition Machine.
93. Özkural, E., *Censored Criticism on the Wikipedia Article About AI Eschatology*, in *log.examachine.net*. 2017.
94. Fagella, D. *Arguments Against Friendly AI and Inevitable Machine Benevolence*. 2020; Available from: https://danfaggella.com/friendly/.
95. Adriano Mannino, D.A., Jonathan Erhardt, Lukas Gloor, Adrian Hutter, Thomas Metzinger. *Artificial Intelligence: Opportunities and Risks*. 2015; Policy paper by the Effective Altruism Foundation (2)]. Available from: https://ea-foundation.org/files/ai-opportunities-and-risks.pdf.
96. Garis, D.H.d., *The Rise of the Artilect Heaven or Hell*.
97. Armstrong, S. *Top 9+2 myths about AI risk*. 2015; Available from: https://www.lesswrong.com/posts/wiwpmtyiKr6bPnSai/top-9-2-myths-about-ai-risk]
98. Hall, B. *superintelligence | Part 4: Irrational Rationality*. Available from: https://www.bretthall.org/superintelligence-4.html.
99. Yampolskiy, R., *On Controllability of AI.* 2020.
100. Manson, M., *I, For One, Welcome Our AI Overlords*, in *Mark Manson*.
101. Häggström, O., *#62 - Häggström on AI Motivations and Risk Denialism*, in *Philosophical Disquisitions*, J. Danaher, Editor. 2019: Philosophical Disquisitions.
102. Bradley, A.J., *Artificial General Intelligence (AGI) is Impeding AI Machine Learning Success*, in *Gartner*. 2019.
103. Schwitzgebel, E., *Against the "Value Alignment" of Future Artificial Intelligence*, in *The Splintered Mind*. 2021.
104. Lauder, E. *AI Won't Seek to Dominate Humanity Says Facebook's Head of AI*. 2017; Available from: https://aibusiness.com/document.asp?doc_id=760243.
105. *Anthropocentrism*, in *Wikipedia*.
106. Naam, R., *What Do AI Researchers Think of the Risks of AI?*, in *Marginal Revolution*. 2015.
107. Häggström, O., *Michael Shermer fails in his attempt to argue that AI is not an existential threat.* Häggström hävdar, 2017.
108. Caplan, B., *My Superintelligence Skepticism*, in *EconLog*. 2013: EconLib.
109. Felten, E. *Why the Singularity is Not a Singularity*. 2018; Available from: https://freedom-to-tinker.com/2018/01/03/why-the-singularity-is-not-a-singularity/.
110. Bokhari, A., *Singularity or Civil War? The Future of Artificial Intelligence*, in *Breitbart*. 2015.
111. Häggström, O., *The AI meeting in Brussels last week*, in *Häggström hävdar*. 2017.
112. Muehlhauser, L. *Three misconceptions in Edge.org's conversation on "The Myth of AI"*. 2014; Available from: https://intelligence.org/2014/11/18/misconceptions-edge-orgs-conversation-myth-ai/.
113. Jeremie Harris, M.M., *Existential risk from AI: A skeptical perspective*, in *Medium*. 2021: towards data science.

114. Jackisch, S., *Superintelligence Skepticism: A Rebuttal to Omohundro's "Basic A.I. Drives"*, in *The Oakland Futurist*. 2016.
115. Anthony Zador, Y.L., *Don't Fear the Terminator*, in *Scientific American*. 2019.
116. Häggström, O., *Johan Norberg is dead wrong about AI risk*, in *Häggström hävdar*. 2018.
117. Chiang, T. *Silicon Valley Is Turning Into Its Own Worst Fear*. 2017; Available from: https://www.buzzfeednews.com/article/tedchiang/the-real-danger-to-civilization-isnt-ai-its-runaway.
118. Schank, R. *HAWKING IS AFRAID OF A.I. WITHOUT HAVING A CLUE WHAT A.I. IS*. 2016; Available from: https://web.archive.org/web/20160531180615/http:/www.rogerschank.com/hawking-is-afraid-of-ai-without-having-a-clue-what-ai-is.
119. Coffee, I., *The most notable luminaries of our time are wrong to fear AI*, in *Medium*. 2016: Input Coffee.
120. Boden, M., *Robot says: Whatever*, in *aeon*. 2018.
121. Goertzel, B., *Superintelligence: Fears, Promises and Potentials | Reflections on Bostrom's Superintelligence, Yudkowsky's From AI to Zombies, and Weaver and Veitas's "Open-Ended Intelligence".* Journal of Evolution and Technology, 2015. **24**(2): p. 55-87.
122. Ballantyne, N., *Epistemic Trespassing.* Mind, 2018. **128**(510): p. 367-395.
123. Steelberg, C. *We need to shift the conversation around AI before Elon Musk dooms us all*. 2017; Available from: https://qz.com/1061404/we-need-to-shift-the-conversation-around-ai-before-elon-musk-dooms-us-all/.
124. Spencer, M., *Elon Musk Is Faking AI Dread for His Neuralink Startup*, in *Last Futurist*. 2020.
125. Tegmark, M. *Elon Musk donates $10M to keep AI beneficial*. 2015; Available from: https://futureoflife.org/fli-projects/elon-musk-donates-10m-to-our-research-program/.
126. Özkural, E., *Scams and Frauds in the Transhumanist Community*, in *log.examachine.net*. 2016.
127. Wheeler, T. *History's message about regulating AI*. 2019; Report Produced by Center for Technology Innovation]. Available from: https://www.brookings.edu/research/historys-message-about-regulating-ai/.
128. Naughton, J., *Don't believe the hype: the media are unwittingly selling us an AI fantasy*, in *The Guardian*. 2019.
129. Ha, A. *Eric Schmidt says Elon Musk is 'exactly wrong' about AI*. 2018; Available from: https://techcrunch.com/2018/05/25/eric-schmidt-musk-exactly-wrong/.
130. Trammell, P., *But Have They Engaged with the Arguments?*, in *Philip Trammell*. 2019. p. #46.
131. Hao, K., *Americans want to regulate AI but don't trust anyone to do it*, in *MIT Technology Review*. 2019.
132. Nicolas Moës, J.B., *AI Alignment Podcast*, in *On the Long-term Importance of Current AI Policy with Nicolas Moës and Jared Brown*, L. Perry, Editor. 2020.
133. felix.h, T.C. *Interview with Tom Chivers: "AI is a plausible existential risk, but it feels as if I'm in Pascal's mugging"*. 2021; Available from: https://forum.effectivealtruism.org/posts/feNJWCo4LbsoKbRon/interview-with-tom-chivers-ai-is-a-plausible-existential.

134. *Friendly AI as a global public good*, in *AI Impacts*.
135. Katja Grace, J.S., Allan Dafoe, Baobao Zhang, Owain Evans, *When Will AI Exceed Human Performance? Evidence from AI Experts.* 2017.
136. Hanson, R., *I Still Don't Get Foom*, in *Overcoming Bias*. 2014.
137. Shah, R. *Coherence arguments do not entail goal-directed behavior*. 2018; Available from: https://www.alignmentforum.org/s/4dHMdK5TLN6xcqtyc/p/NxF5G6CJiof6cemTw.
138. O'Sullivan, L. *I Quit My Job to Protest My Company's Work on Building Killer Robots*. 2019; ACLU]. Available from: https://www.aclu.org/blog/national-security/targeted-killing/i-quit-my-job-protest-my-companys-work-building-killer.
139. Cassauwers, T. *CONFUCIANISM COULD PUT FEARS ABOUT ARTIFICIAL INTELLIGENCE TO BED*. 2019; "The Daily Dose"]. Available from: https://www.ozy.com/the-new-and-the-next/how-confucianism-could-put-fears-about-artificial-intelligence-to-bed/93206/.